\documentclass[superscriptaddress,preprint,aps,onecolumn]{revtex4-2}
\usepackage[utf8]{inputenc}
\usepackage{mathtools}
\usepackage{bm}
\usepackage{gensymb} % degree symbol and some others
\usepackage{graphicx} % Required for the inclusion of images

\usepackage{hyperref}
\hypersetup{colorlinks,
    citecolor=black,
    filecolor=black,
    linkcolor=blue,
    urlcolor=black
}
\usepackage{tabularx}
\newcolumntype{L}{>{$}l<{$}} % math-mode version of "l" column type
\newcolumntype{R}{>{\raggedleft\arraybackslash}X}
\newcommand{\rev}[1]{\textcolor{black}{#1}}

\usepackage{subfigure}
\usepackage{url}
\usepackage{float}
\usepackage{color}

\begin{document}
\title{Quantum Nuclei at Weakly Bonded Interfaces: The Case of Cyclohexane on Rh(111)}

\author{Karen Fidanyan}
\affiliation{Fritz Haber Institute of the Max Planck Society, Faradayweg 4-6, 14195 Berlin, Germany}
\affiliation{Max Planck Institute for the Structure and Dynamics of Matter, Luruper Chaussee 149, 22761 Hamburg, Germany}

\author{Ikutaro Hamada}
\email{ihamada@prec.eng.osaka-u.ac.jp}
\affiliation{Department of Precision Engineering, Graduate School of Engineering, Osaka University, 2-1 Yamadaoka, Suita, Osaka 565-0871, Japan}

\author{Mariana Rossi}
\email{mariana.rossi@mpsd.mpg.de}
\affiliation{Fritz Haber Institute of the Max Planck Society, Faradayweg 4-6, 14195 Berlin, Germany}
\affiliation{Max Planck Institute for the Structure and Dynamics of Matter, Luruper Chaussee 149, 22761 Hamburg, Germany}

\begin{abstract}
The electronic properties of interfaces can
depend on their isotopic constitution. One known case is that of 
cyclohexane physisorbed on Rh(111), in which isotope effects have been measured on the work function change and desorption energies. These effects can only be captured by calculations including nuclear quantum effects (NQE). 
In this paper, this interface is addressed employing dispersion-inclusive density-functional theory coupled to a quasi-harmonic (QH) approximation for NQE, as well as to fully anharmonic \textit{ab initio} path integral molecular dynamics (PIMD).
The QH approximation is able to capture that deuterated cyclohexane has a smaller adsorption energy and lies about 0.01 \r{A} farther from the Rh(111) surface than its isotopologue, which can be correlated to the isotope effect in the work function change. 
An investigation of the validity of the QH approximation relying on PIMD simulations, leads to the conclusion that although this interface is highly impacted by anharmonic quantum fluctuations in the molecular layer and at bonding sites,
these anharmonic contributions play a minor role when analysing isotope effects at low temperatures. Nevertheless, anharmonic quantum fluctuations cause an increase in the distance between the molecular layer and Rh(111), a consequent smaller overall work function change, and intricate changes in orbital hybridization.
\end{abstract}

\maketitle

\section{Introduction}
Usually, the electronic properties of interfaces do not strongly depend on the isotopic constitution of the atoms that compose them. This is because the electronic structure of different isotopes is the same and nuclei can typically be considered as classical particles, which means that an isotopic change cannot lead to a change in the (static) atomic structural properties of materials, and thus do not cause a change in the electronic structure. 
However, when the quantum nature of the nuclei makes itself more prominent, this ceases to be true. An isotopic change can lead to structural changes of the material and thus to a  considerable change in the electronic structure. 
Such electron-phonon coupling effects can be captured to a great extent in the adiabatic limit~\cite{Giustino2017}. In this case,  electronic properties can be modified because of their dependence on the nuclear positions and the equilibrium distribution of nuclear fluctuations at any given temperature.

One known case to exhibit such isotopic effects is cyclohexane ($\rm C_6H_{12}$) adsorbed on platinum-group metal surfaces. It was shown in a series of papers by Koitaya, Yoshinobu, and coworkers~\cite{Koitaya_2012,Koitaya_2014}, that the change of work function induced by adsorbed cyclohexane is different when considering $\rm C_6H_{12}$ and fully deuterated $\rm C_6D_{12}$. 
Based on work function measurements and previous calculation of alkanes on metal surfaces~\cite{Morikawa_2004}, it was suggested that deuterated molecules should lie farther from the surface.
Also the desorption energy differs significantly: that of $\rm C_6H_{12}$ on Rh(111) is \rev{84 $\pm$ 23 meV higher than that of $\rm C_6D_{12}$ at lower coverages}, thus showing an inverse kinetic isotope effect. % on the desorption rate. 
In these systems, such effects are of a certain relevance because changes in the strength of the bond between hydrogen and metal and between hydrogen and carbon impacts the dehydrogenation propensity of cyclohexane -- a molecule that often plays a central role in systems aiming at cheap high-density hydrogen storage \cite{Li:2015iz}. The availability of experimental data and the importance of these systems thus make them an ideal ground to study the performance of different theoretical techniques in a complex but well-defined environment.

The challenges for theory to tackle this problem stem from the necessity of capturing complex electronic-structure changes, as well as the multidimensional atomic structure of quantum nuclei. In particular, electron-phonon coupling needs to be included at least in an approximate fashion in order to relate nuclear fluctuations and electronic-structure variations.
Modelling these effects becomes more important as the field moves towards soft and hybrid electronic materials, where electron-phonon coupling tends to be more pronounced~\cite{Koch_2007, Jacobs_Wang_2020}.
A common way to address such problems is to employ the harmonic approximation for the nuclear vibrations on first-principles potential energy surfaces~\cite{Patrick_Guistino_2014}.
However, the validity of this approximation in weakly bonded systems and interfaces, where anharmonic terms in the potential energy surface are expected to play a role, is questionable.

Instead, a method capable of including NQE without relying on the harmonic approximation is \textit{ab initio} path integral molecular dynamics (aiPIMD)~\cite{Parrinello_Rahman1984}. 
Despite its immense potential, a significant drawback of aiPIMD simulations is their high computational cost. Therefore, in this work  aiPIMD simulations are performed making use of a technique that reduces the amount of replicas required for simulations of weakly-bonded interfaces~\cite{SL-RPC}.
These results are compared to harmonic and quasi-harmonic approximations. With these simulations, we are able to explain the physical origin of the observed isotope effects on the cyclohexane/Rh(111) interface and identify when a quasi-harmonic analysis of these effects is valid.
\rev{We study the impact of nuclear quantum fluctuations on the electronic structure taking advantadge of the aiPIMD simulations -- an approach that has been successfully applied previously on diverse systems~\cite{Kaczmarek_etal_2009, Chen_etal_Pasquarello_2016, Law_Hassanali_2015}.}
The capabilities and limitations of aiPIMD based on density-functional theory are further discussed \rev{for this system}.

\section{Results and discussion}

\begin{figure}[ht]
    \begin{center}
        \includegraphics[width=0.49\textwidth]{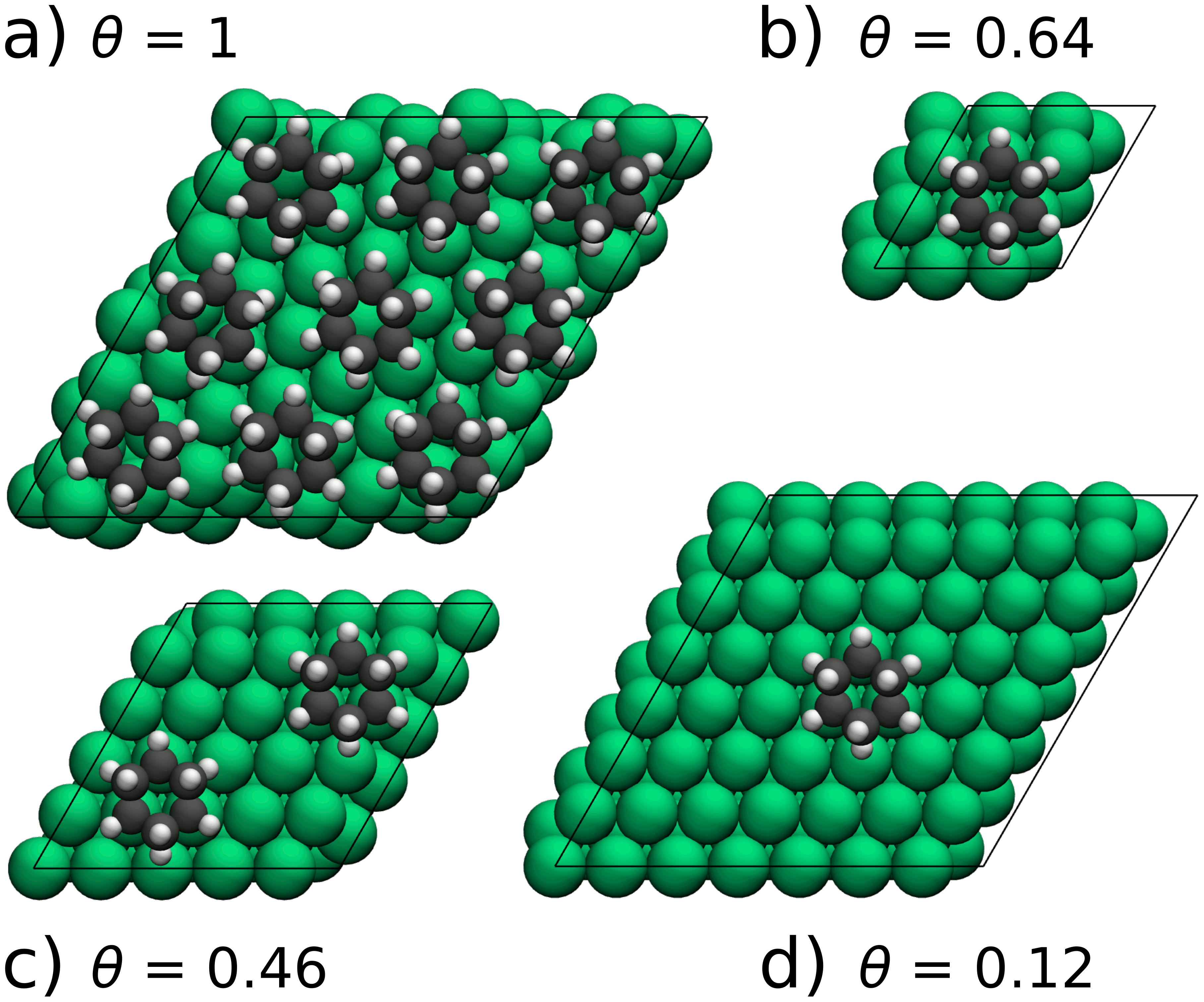}
        \caption{The cyclohexane adsorption patterns considered in this work for modelling different converages $\theta$.
        a) $\theta = 1$,  $(2\sqrt{3}\times 2\sqrt{3})R13.9\degree$ unit cell \rev{(0.173 molecules per Rh atom)}.
        b) $\theta = 0.64$,  $(3 \times 3)$ unit cell. 
        c) $\theta = 0.46$,  $(5 \times 5)$ unit cell.
        d) Coverage $\theta = 0.12$,  $(7 \times 7)$ unit cell.
            } \label{fig_commensurate}
    \end{center}
\end{figure}

Experimental measurements conducted on the cyclohexane/Rh(111) interface have shown that several aspects of the adsorption show a dependence on the coverage~\cite{Koitaya_2012}. In particular, desorption competes with dehydrogenation at coverage values below 0.5. 
We therefore built models for the different coverages, based on existing experimental data.
On a clean Rh(111) surface, experiment shows a high-order large commensurate $(2\sqrt{79} \times 2\sqrt{79})R17.0\degree$ pattern~\cite{Koitaya_2011}.
This system size would not be computationally tractable, given the amount of \textit{ab initio} simulations necessary to investigate nuclear quantum effects in this system.
A smaller commensurate structure that was also observed in experiment is a $(2\sqrt{3}\times 2\sqrt{3})R13.9\degree$ pattern~\cite{Koitaya_2013}, shown in figure~\ref{fig_commensurate}a. This structure was taken as a reference for the full-coverage monolayer structure.
The effective coverage for other structures derived from this one were calculated, and smaller unit cells for lower coverage were modelled. 
These models are shown in figure~\ref{fig_commensurate}b-d.
In the following, \rev{we perform our calculations principally on the structure presented in figure~\ref{fig_commensurate}c ($\theta = 0.46$) unless explicitly stated otherwise. This supercell, containing two cyclohexane molecules and a Rh(111) ($5\times5$) surface cell, allows us to capture most of the phonon band structure dispersion, which is especially pronounced for the metal surface~\cite{BOHNEN1996222}, in the real-space dynamics simulations.}

\subsection{Static results and the quasi-harmonic approximation}

\begin{figure*}[ht]
    a)\\
    \begin{center}
    \begin{tabular}{|r|r|r|r|r|r|r|r|}
    \hline
    coverage $\theta$ & $E^{\rm{pot}}_{\rm{ads}}$ & 
     $E^{\rm{pot}}_{\rm{ads}}$+ZPE(H) & $E^{\rm{pot}}_{\rm{ads}}$+ZPE(D) & $\Delta$ ZPE & $F^{\rm{harm}}_{\rm{ads}}(\rm{H})$ & $F^{\rm{harm}}_{\rm{ads}}(\rm{D})$ & $\Delta F(\rm{H}-\rm{D})$ \\ \hline
    0.12 & 945  & 1039 & 1004 & 35 & 742 & 705 & 37 \\ 
    0.46 & 953  & 1056 & 1022 & 34 & 786 & 750 & 36 \\ 
    0.64 & 946  & 1046 & 1013 & 33 & 780 & 745 & 35 \\ 
    1.0  & 1023 & 1066 & 1049 & 17 & 790 & 770 & 20 \\ 
    \hline
    0.3 (TPD~\cite{Koitaya_2012}) &  &   728 $\pm$ 12 & 644 $\pm$ 20 & 84 $\pm$ 23 &  &   &  \\ \hline 
    \end{tabular}
    \end{center}
    
    b)
    \begin{center}
    \includegraphics[width=0.5\textwidth]{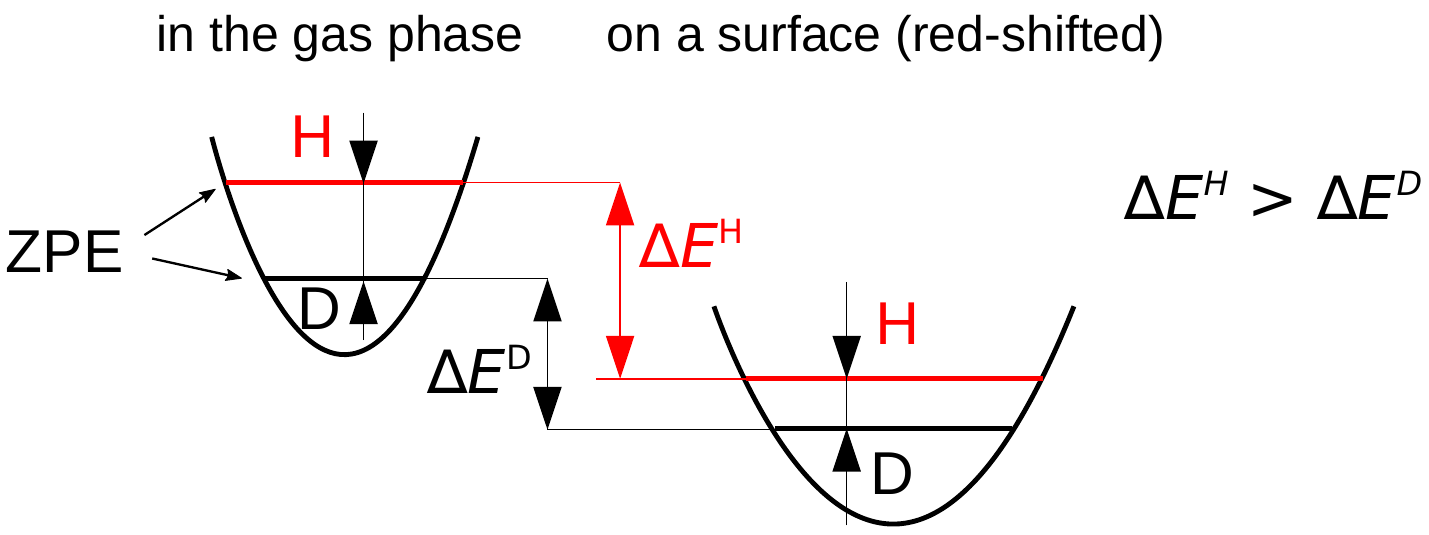}
    \caption {a)  Adsorption energies and harmonic free energies for different coverage values, calculated with the PBE+vdW$^{\text{surf}}$ functional (\textit{light} settings) \rev{according to eq.~\ref{eq_Eads}}. The free energy is calculated for the temperature of 150~K and all energies are in meV. Experimental data from temperature programmed desorption (TPD) experiments from Ref.~\cite{Koitaya_2012}.
    b) The effect of the red shift in the C-H stretching modes on the adsorption energy, shown schematically. The difference in ZPE is between H and D is higher in vacuum than on surface, due to the different masses and the red-shift of the corresponding stretching mode upon binding.
    }\label{fig_sketch}
    \end{center}
\end{figure*}

For the different coverages shown in figure~\ref{fig_commensurate}, the adsorption energy $E_{\text{ads}}^{\text{pot}}$ per molecule was calculated as explained in Methods.
The results employing the PBE+vdW$^{\text{surf}}$\rev{~\cite{TS2009, Ruiz2012}} functional 
are reported in \rev{column 2 of figure~\ref{fig_sketch}a}. \rev{Comparison between \textit{Light} and \textit{Tight} computational settings is given in table~S1 in Supplementary Information (SI)}.

The harmonic free energy terms at a temperature of 150 K was added to the adsorption energy, as given by eq.~\ref{eq_Fharm}. 
This temperature was chosen to satisfy several conditions.
On one hand, the temperature should be below the temperature of desorption and dehydrogenation, which are both close to 200~K~\cite{Koitaya_2012}. 
On the other hand, very low temperatures considerably increase the cost of a PIMD simulation and would reduce the role of anharmonicity, which we aim to investigate.
The adsorption free energies $F^{\rm{harm}}_{\rm{ads}}$ were obtained analogously to 
eq.~\ref{eq_Eads}.

The results for each coverage are summarized in  figure~\ref{fig_sketch}a, columns \rev{3-8}.
The addition of the zero point energy and the temperature-dependent free energy terms, already in the harmonic approximation, lead to a different adsorption energy for $\rm C_6H_{12}$ and $\rm C_6D_{12}$. 
This is to be expected, because the C-H stretching modes of the adsorbed cyclohexane associated with the CH groups that point to the surface show a significant red shift of up to 300 cm$^{-1}$ in comparison to the gas phase, as shown in figure~S2 in SI. 
Because of the difference in mass between the H and D atoms, such a red shift has a stronger impact on the ZPE of a \rev{C-H} vibration, compared to a \rev{C-D} one, as schematically shown in figure~\ref{fig_sketch}b.
In both cases, the effect of ZPE increases the energy of adsorption (figure~\ref{fig_sketch}a, columns \rev{3,4}), and in the case of $\rm C_6H_{12}$ this effect is stronger. When adding the full free energy contributions, the translational and rotational entropic contributions of the gas-phase molecules work to decrease the adsorption free energy (figure~\ref{fig_sketch}a, columns \rev{6,7}). The magnitude of the isotope effect is not strongly affected by temperature and is about a factor two smaller than what is observed in \rev{temperature programmed desorption} experiments\rev{~\cite{Koitaya_2012}, as shown in figure~\ref{fig_sketch}a}.
We also observe that the difference between H/D adsorption energies becomes smaller at full coverage. It decreases from 37 meV for $\theta = 0.12$ down to 20 meV for $\theta = 1$. 
This \rev{trend appears} because the red shift of surface-pointing C-H stretch modes decreases with increasing coverage, pointing to a weaker molecule-surface interaction (see SI, figure~S2). This confirms the weakening of the Rh-H bond with increasing coverage (and consequent strengthening of the C-H bond).

\begin{figure}[htbp]
    \begin{center}
        \includegraphics[width=0.48\textwidth]{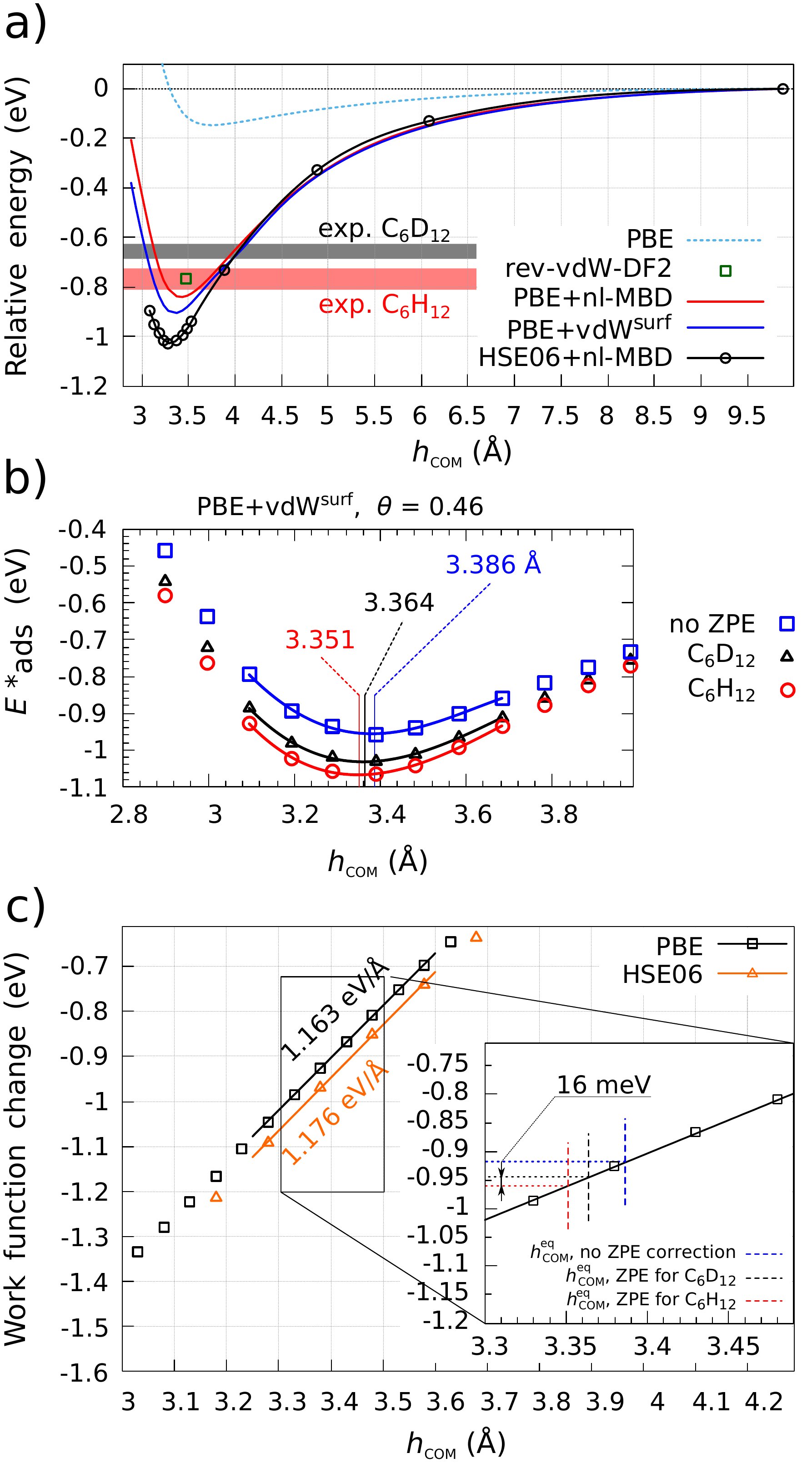}
        \caption{a) Adsorption curves calculated with different exchange-correlation functionals and vdW corrections: PBE (dotted blue line), PBE +$\rm{vdW^{surf}}$ (solid blue line), PBE+nl-MBD (solid red line), \rev{HSE06+nl-MBD (black points)}, and rev-vdW-DF2 (green square, equilibrium distance only). Calculations were performed with the unit cell of $\theta = 0.46$.
        Shaded areas show the \rev{interval of reported} experimental values of the adsorption energy of $\rm{C_6H_{12}}$ (red) and  $\rm{C_6D_{12}}$ (grey) \rev{around the coverages we study}~\cite{Koitaya_2012}.
        \rev{b) ZPE-corrected energy of adsorption for $\rm{C_6H_{12}}$ (red) and $\rm{C_6D_{12}}$ (black), calculated according to eq.~\ref{eq:quasiharm-eads} with PBE + $\rm{vdW^{surf}}$. Blue line shows the adsorption energy values calculated without ZPE correction.
        c) Work function change as a function of distance to surface, calculated with PBE (black squares) and HSE06 (orange triangles)}. Vertical dashed lines in the inset mark the equilibrium distances for classical nuclei (blue), $\rm{C_6D_{12}}$ (black) and $\rm{C_6H_{12}}$ (red).
        } 
        \label{fig_static_data}
    \end{center}
\end{figure}

In order to \rev{investigate} the impact of the exchange-correlation functional on the \rev{description of this system, we have computed the adsorption profiles of cyclohexane on Rh(111) with different functionals and vdW corrections}.
In figure~\ref{fig_static_data}(a), we show the adsorption curve with the PBE functional, the PBE+vdW$^{\text{surf}}$ functional, the PBE functional with the recently proposed many-body dispersion method nl-MBD~\cite{mbd_nl_2020}, \rev{the range-separated hybrid functional HSE06~\cite{hse06} combined with nl-MBD dispersion interactions}, and the non-local rev-vdW-DF2~\cite{Hamada_rev_vdW_DF} functional (see Methods for the details of calculations), \rev{all for the coverage of $\theta=0.46$}. \rev{The minima of these curves are tabulated in table~S2 in SI. }
\rev{The rev-vdW-DF2 calculation was performed only at the equilibrium position and at the reference geometries (i.e., isolated slab and molecule). Further calculations of the binding curve with this functional for a different coverage ($\theta=0.64$) are reported in the SI section~3 and table~S3.}

\rev{We start by analyzing the dataset based on the PBE functional, which allows us to understand the impact of different descriptions of the vdW corrections. Comparing the result obtained with the bare PBE functional and the others, we conclude, as expected, that}
vdW interactions are a fundamental piece of the molecule-surface interaction. 
\rev{We then proceed to compare the results obtained with PBE+vdW$^{\text{surf}}$ with the results obtained with  PBE+nl-MBD. Both of these vdW corrections do not enter the Kohn-Sham potential within the self-consistent procedure, such that they cannot change the electronic density. 
We observe that vdW$^{\text{surf}}$ predicts a larger binding energy (0.90 eV) and an equilibrium distance closer to the surface (3.36 \AA) than nl-MBD (0.84 eV and 3.45 \AA, respectively).
Considering that nl-MBD  contains explicit many-body vdW effects and captures the electronic screening of these interactions better than vdW$^{\text{surf}}$~\cite{mbd_nl_2020}, we can conclude that the observed differences are due to both of these effects. 
Then, we can compare the results obtained with the PBE+nl-MBD and the HSE06+nl-MBD functionals. In this case, the vdW interactions are treated at the same level, but the short-range exchange term is modified to include a fraction of exact exchange. 
The changes in the electronic density brought by the HSE06 functional produce a larger binding energy (1.03 eV) and an equilibrium position that is even closer to the surface (3.30 \AA). 
These observations, combined, point to the prediction of stronger bonds between surface and adsorbate when the self-interaction error is mitigated.
Finally, the rev-vdW-DF2 functional, which improves the description of the polarizability of these systems, predicts the adsorption distance of 3.48 \r{A} and binding energy of 0.77 eV, which is smaller than those obtained with other functionals. The experimental values from TPD experiments lie closer to the PBE+nl-MBD and rev-vdW-DF2 binding energies.}
For further investigation, we employ \textit{light} settings of the FHI-aims code and a comparison of the adsorption curve between \textit{tight} and \textit{light} settings is shown in figure~S1 in SI.
The shape of the potential along the desorption coordinate hints that anharmonic effects on this coordinate or others that couple with it could have an important effect on the adsorption properties of this system, and potentially on nuclear quantum effects. 

In order to investigate the impact of nuclear quantum effects including anharmonicity at least on the desorption coordinate, harmonic phonons and the corresponding zero-point-energy contribution to the adsorption energy for the adsorbed $\rm{C_6H(D)_{12}}$ at different distances to Rh surface were calculated. 
At each desired distance, we have fixed the center of mass of the molecule, and \rev{optimized the other degrees of freedom except the two bottom layers}. 
See the discussion in the SI \rev{section~4} regarding the inclusion of different vibrational modes in this ZPE correction. The quasi-harmonic (QH) ZPE-corrected energy of adsorption $E^{*}_{\rm{ads}}$ was then calculated according to eq. \ref{eq:quasiharm-eads}.
These values were calculated with the PBE+vdW$^{\rm{surf}}$ functional, \rev{because it delivers a good description of the desorption curve, and its computational cost and implementation in the FHI-aims code allow a fast computation of thousands of force evaluations -- which is not the case for the other functionals we show here.} 
The results are presented in figure \ref{fig_static_data}b.
The adsorption energies and distances obtained for $\rm{C_6H_{12}}$ and $\rm{C_6D_{12}}$ in this way are also summarized in table~S5 in the SI. 
With this procedure, a deformation of the binding energy curve that is different for $\rm{C_6H_{12}}$ and $\rm{C_6D_{12}}$ is predicted, such that $\rm{C_6H_{12}}$ has a larger binding energy and adsorbs closer to the surface than $\rm{C_6D_{12}}$.
The H-D adsorption energy difference is \rev{\textbf{37} meV} for PBE + $\rm{vdW^{surf}}$ functional, \rev{which is slightly larger than the value reported in figure~\ref{fig_sketch}a.} 
\rev{In the SI (figure~S3) we show a comparison of this QH procedure with the rev-vdW-DF2 functional and a different coverage, which confirms that the effect of ZPE on these curves is quite similar across different coverages and functionals.}
We have checked that adding finite temperature contributions in the harmonic approximation to these values, up to 150 K, does not appreciably change this calculated isotope effect either (see SI, figure~S4).

\rev{Regarding the equilibrium distance of absorption, there is an important effect that is observed.
The adsorption distance of $\rm{C_6H_{12}}$ is \rev{\textbf{3.351}} \r{A}, and for $\rm{C_6D_{12}}$, it is \rev{\textbf{3.364}} \r{A}.} % for PBE-$\rm{vdW^{surf}}$ (rev-vdW-DF2) calculations.
The H-D equilibrium distance difference is thus $\approx$\textbf{0.01} \r{A}. Although apparently small, this equilibrium distance difference has a \rev{marked} effect on the work function changes.
The sensitivity of the work function change $\Delta \phi$ of the interface to the distance between the adsorbate and the metal surface is shown in figure~\ref{fig_static_data}c. We calculated it by shifting an adsorbate rigidly closer and farther from the slab with respect to the equilibrium position at the potential energy surface. Around the equilibrium distance, the work function depends almost linearly on the distance, and the slope is of \rev{\textbf{1.16} eV/\AA~ with the PBE functional and of \textbf{1.18} eV/\AA~ with HSE06. The $\Delta \phi$ with HSE06 is about 0.06 eV larger than with the PBE functional.}
A change of adsorption distance as the one observed between the deuterated and normal cyclohexane (0.01~\r{A}) would thus \rev{result in an isotope effect on} $\Delta \phi$ of \textbf{16-17} meV, which does not strongly depend on the functional. In experiment~\cite{Koitaya_2012}, the same qualitative trend was observed, but a \rev{slightly} larger isotope effect on  $\Delta \phi$ was reported, namely of \rev{$\approx$ 25~meV} at $\theta = 0.46$. \rev{This value was obtained by a linear regression of the experimental data for the $\Delta \phi$ dependence on coverage reported in Ref.~\cite{Koitaya_2012}, in the interval $0.1 < \theta < 0.65$, followed by an alignment of the fits such that they yield $\Delta \phi = 0$ at $\theta$=0. In the SI, figure~S5, we show that larger coverages considerably increase the value and the slope of $\Delta \phi$ with distance from the surface, but the inclusion of self-consistent vdW interactions does not change appreciably the values or the slopes of $\Delta \phi$.}

\begin{figure*}[htbp]
    \begin{center}
        \includegraphics[width=0.9\textwidth]{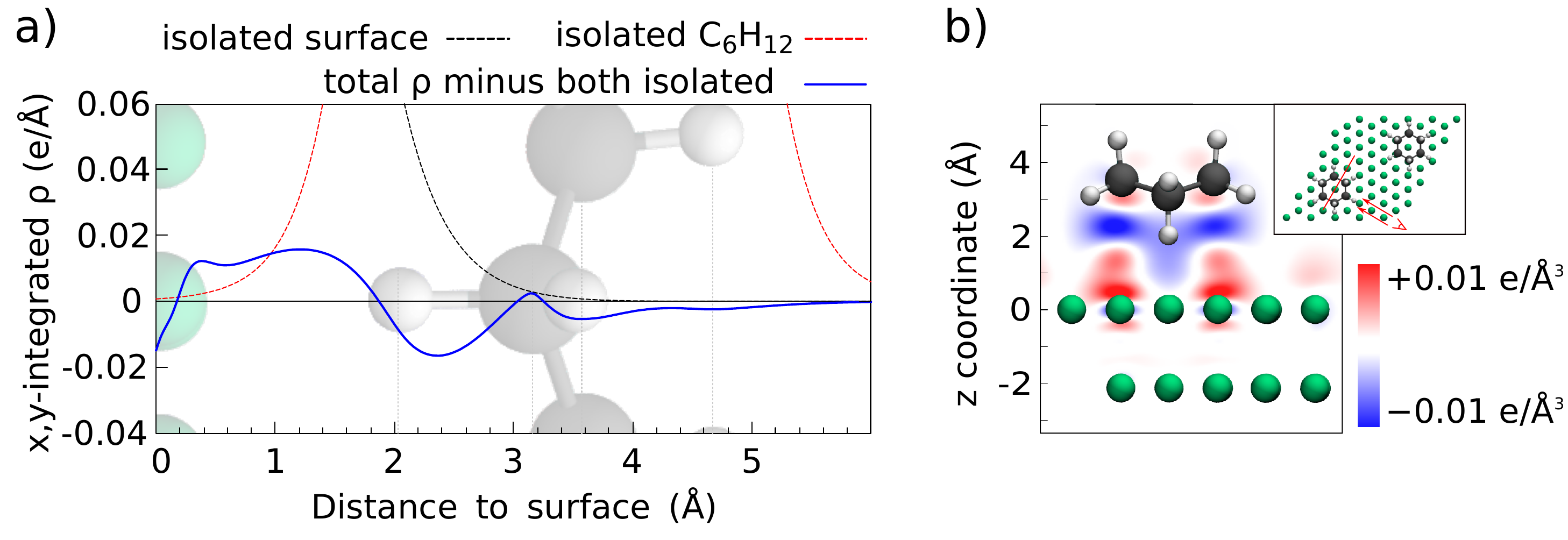}
        \caption{a) The electron density $\rho$ of the interface integrated over the $x$ and $y$ dimensions and projected on the $z$ axis. The dashed lines represent the electron densities of the clean surface and the adsorbate monolayer, calculated separately. The blue line shows difference between the total density and the superposition of isolated parts. It shows the density accumulation between Rh and H atoms and the depletion in C-H bond.
        b) The difference between the total electron density of the interface and the sum of the densities of the clean surface and the isolated adsorbate. Blue shows electron depletion, and red shows accumulation. The inset shows the unit cell and the slicing plane (the dashed red line). The position of the plane is chosen as shown in the inset.}
        %in the same way as in the figure~4 in~\cite{Bagus_Hermann_Woll_2005}.}
        \label{fig_rho}
    \end{center}
\end{figure*}

\rev{The origin of the work-function change upon the adsorption of a neutral and non-polar adsorbate like cyclohexane,} is typically attributed (i) to the so-called ``pushback effect'', when molecules ``push'' the vacuum tail of the electron density of the metal back into a surface~\cite{Ishii_et_al_1999, Bagus_Hermann_Woll_2005}, and (ii) to the polarization of the adsorbate induced by the mirror image charge formed in metal~\cite{Wandelt_Xe_1984}. 
By analyzing electronic charge density differences from PBE, shown in figure~\ref{fig_rho}, we observe the pushback effect but also \rev{density rearrangements that are typical of weak bond formation.}
To visualize this, the electronic densities were integrated over the directions parallel to the surface and projected in the perpendicular direction (figure~\ref{fig_rho}a). The total electronic density of the interface was compared with the sum of the electronic densities of the clean surface and the isolated adsorbate. 
We observe an electron depletion in the C-H bonds and accumulation in the H...Rh region, which is associated with H-Rh bond formation.
It is accompanied by the pushback effect similar to what was reported by Bagus \textit{et al.} for cyclohexane on a Cu(111) cluster~\cite{Bagus_Hermann_Woll_2005}. In figure~\ref{fig_rho}b, the redistribution of the electron density on a plane perpendicular to the surface, that crosses only carbon-carbon bonds, is shown. In this plane, the contribution of \rev{H-metal bonds} is small, therefore the electron depletion that is present under the molecule can be attributed to a pushback effect.
It is difficult to separate the role of \rev{H-metal bond} formation and the pushback effect clearly. 
By investigating the spatial arrangement of the electronic density changes, we conclude that the contributions of these two effects are of comparable magnitudes (see SI, figures S6 and S7).
Together, \rev{H-metal bonds} and a pushback effect cause a \textbf{2.13} Debye per molecule decrease in the dipole moment of the interface.

\subsection{Validity of the quasi-harmonic analysis} \label{section_validity_of_HA}

\begin{figure*}[htbp]
    \begin{center}
        \includegraphics[width=0.94\textwidth]{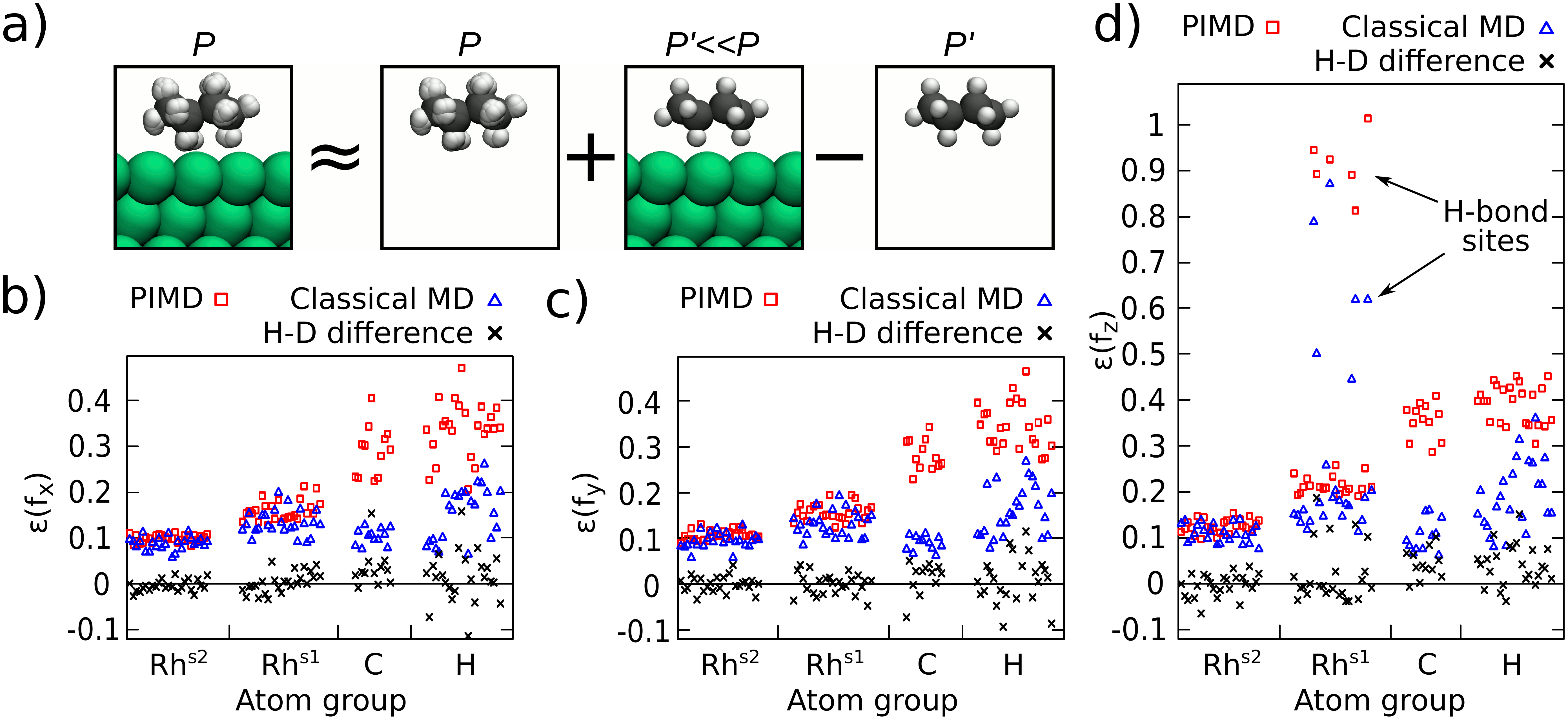}
        \caption{a) A scheme of the spatially-localized ring polymer contraction (SL-RPC). The forces for a full ring polymer of $P$ beads are approximated by a superposition of forces calculated for $P$ beads at the molecular part, $P'<P$ beads of the full system and a correction of $P'$ beads at the molecular part.
        b,c) Anharmonicity measure $\epsilon$ (see eq. \ref{eq:anh-score}) for individual Cartesian $x$ and $y$ components of atomic forces from the PIMD simulation of $\rm{C_6H_{12}}$ (red squares) compared to classical-nuclei MD (blue triangles), and difference in $\epsilon$ between PIMD simulations of $\rm{C_6H_{12}}$ and $\rm{C_6D_{12}}$ (black crosses). All values calculated for $\theta = 0.46$ (two cyclohexane molecules in the unit cell) and at $T = 150$ K. $\rm{Rh^{s1}}$ and $\rm{Rh^{s2}}$ denote the 1st and the 2nd layers of the surface atoms.
        d) Anharmonicity measure $\epsilon$ for the Cartesian $z$ component of the forces. The distinct group of Rh atoms with highly anharmonic forces consists of atoms connected to cyclohexane via hydrogen-metal bonds.
        } \label{fig_epsilon}
    \end{center}
\end{figure*}

The quasi-harmonic treatment and the relationship between the distance of the molecule to the surface and the work function change already provide a qualitative explanation about the electronic structure changes that take place upon isotopic substitution. At this point, it is interesting to gauge the accuracy the quasi-harmonic approximation in these types of interfaces.

Path integral molecular dynamics (PIMD), a method which exploits the quantum-classical isomorphism between a quantum system and a classical ring polymer with infinite number of beads $P$~\cite{Parrinello_Rahman1984}, can be employed in this context, in order to calculate static thermodynamic averages without relying on any harmonic ansatz.
When performed on \textit{ab initio} potentials (aiPIMD), subsequent analysis can capture the coupling between quantum molecular vibrations and the electronic structure of a surface in the adiabatic limit, with full anharmonicity and full dimensionality.

However, given the need of a first-principles potential energy surface and the large system sizes involved, 
aiPIMD simulations of this sort are associated with a high simulation cost, because of the need of several replicas of the full system to obtain converged results (the number of beads $P$, which controls the convergence of PIMD results to exact quantum expectation values). In order to alleviate this cost, one can use the fact that the bonding between molecules and a surface is relatively weak in the case of cyclohexane,  to apply the spatially-localized ring-polymer contraction (SL-RPC)
~\cite{SL-RPC}. The core idea of this contraction is sketched in figure~\ref{fig_distributions}a. The potential energy (and the corresponding forces) of the full system of $P$ beads is approximated as a superposition of $P$ replicas of the molecular part, $P' \ll P$ replicas of the full system, and additional $P'$ of the molecular part with negative sign, i.e.
\begin{equation}
\begin{split}
    V_P(\bm{q}) \approx & 
    \frac{P}{P'} \sum_{k=1}^{P'} \left[ V_{\rm{full}}(\bm{\tilde{ q}}_{\rm{full}}^{(k)}) - V_{\rm{mol}}(\bm{\tilde{ q}}_{\rm{mol}}^{(k)}) \right] \\
    & + \sum_{k=1}^P V_{\rm{mol}}(\bm{q}_{\rm{mol}}^{(k)}), \label{eq:pot-approx}
\end{split}
\end{equation}
where ``full'' denotes the full system containing the surface and adsorbed molecules, and ``mol'' denotes the adsorbate simulated in the same unit cell, but without the surface. $\bm{q^{(k)}}$ are Cartesian positions of beads, and $\bm{\tilde{q}^{(k)}}$ are obtained by the Fourier interpolation of the full
$P$ beads ring polymer to the contracted $P'$-polymer.

This approximation, of course, does not come without errors. One can estimate the error introduced by SL-RPC in a harmonic potential~\cite{SL-RPC}
\begin{equation}
\begin{split}
    \delta E^{\rm{RPC}} & = E^{\rm{RPC}} - E^{\rm{P\ beads}} = \\
    & = \sum_{i=1}^{3N} \frac{k_B T}{2} \sum_{k=P'}^{P-1} \left[ 
    \frac{\omega^2_{\rm{mol}}}{\omega_k^2 + \omega^2_{i,\rm{mol}}} - 
    \frac{\omega^2_{\rm{full}}}{\omega_k^2 + \omega^2_{i,\rm{full}}} \right] , \label{eq_SLRPC_E_error}
\end{split}
\end{equation}
where $\{\omega_{i,\rm{mol}}\}$ are normal modes of the isolated adsorbate, and $\{\omega_{i,\rm{full}}\}$ are the corresponding modes calculated by diagonalization of the part of Hessian matrix that describes molecular displacements. Equation \ref{eq_SLRPC_E_error} is slightly different from eq. 9 in Ref.~\cite{SL-RPC} because we have not made the assumption of $\omega^2_{\nu,\rm{mol}} - \omega^2_{\nu,\rm{full}} \ll \omega_k^2 + \omega^2_{\nu,\rm{full}}$, as discussed in section~7 in the SI. 

Similarly, one can estimate the error in a harmonic quantum free energy at finite temperatures (see the derivation in the SI, section~8) as,
\begin{equation}
    \delta F^{RPC} = \frac{1}{2\beta} \sum_{i = 1}^{3N} \sum_{k=P'}^{P-1} \ln\left(1 + \frac{\omega^2_{i,\rm{full}} - \omega^2_{i,\rm{mol}}}{\omega_k^2 + \omega^2_{i,\rm{mol}}} \right).
    \label{eq_SLRPC_F_error}
\end{equation}

Such an estimate for a cyclohexane on Rh(111), when taking $P'=1$ does not exceed \textbf{37}~meV per molecule for the total potential energy and \textbf{79}~meV/molecule for the total free energy. When comparing H- and D-cyclohexane, one can rely on error cancellation. Then, the error in potential energy \textit{difference} is \textbf{19} meV/molecule, and in free energy difference about \textbf{36} meV/molecule. It is thus clear that although this approximation is very useful, if quantitative results for this particular system are desired, a contraction to the centroid ($P'=1)$ is not sufficient and we do not further pursue calculations of free energies at this level of approximation. It should, however, be sufficient to capture further important anharmonic effects if present.

Before analyzing the PIMD results, we study the anharmonic contributions to the forces in this system, separating them into classical finite temperature effects and nuclear quantum effects. For this purpose, we follow the lines of  Ref.~\cite{Knoop_2020} and calculate an anharmonicity measure for different degrees of freedom. Because it is interesting to compare the difference between quantum and classical anharmonic contributions to different coordinates in the system, we calculate
\begin{equation}
    \epsilon(T)^{\rm{CL/QM}} = \sqrt{\frac{\langle (F_{\rm{DFT}}^{\rm{CL/QM}} - F_{\rm{h}}^{\rm{CL/QM}})^2 \rangle_T} {\sigma^2_{F_{\rm{DFT}}^{\rm{QM}}}(T)}}, \label{eq:anh-score}
\end{equation}
where $F_{\rm{DFT}}$ are the full DFT forces calculated, $F_{\rm{h}}$ are harmonic forces calculated for the same geometry and with the Hessian matrix obtained for the full system, $\sigma^2$ is the variance, and $\langle \dots \rangle_T$ is the ensemble average at the temperature $T$. The superscripts CL and QM denote a classical-nuclei (aiMD) and a quantum-nuclei (aiPIMD) simulations, respectively. In the latter case, we take the forces on the bead positions. 
 We normalize both classical and quantum quantities by the respective PIMD variance of that quantity so that the difference $\epsilon(T)^{\rm{QM}} - \epsilon(T)^{\rm{CL}}$ corresponds to a measure of the amount of ``quantum anharmonicity''.
 
We show the results of this analysis for $\rm{C_6H_{12}}$ and $\rm{C_6D_{12}}$ in figure~\ref{fig_epsilon}~{(b-d)}, where the force components are resolved into the three Cartesian directions (all calculations with PBE+vdW$^{\text{surf}}$).
Such an analysis shows, as expected, that anharmonic contributions are more pronounced in the adsorbate molecules, and the difference between quantum and classical anharmonic scores is really only pronounced on the molecular adsorbate. However, H-bonded sites on the top layer of the surface not only show a pronounced anharmonicity in the $z$ direction, but also a considerable quantum component, no doubt caused by the nuclear quantum contributions to the \rev{H-metal bond}. In fact, comparing the $z$ component of the forces with the $x$ and $y$ components, the $z$ anharmonic score is always higher for the top surface layer and the molecules. This supports the conclusion that most anharmonic effects lie on the out-of-plane motions of molecules, which are captured to a limited extent in the QH approximation. However, these calculations also show that the contribution of quantum anharmonicity on $\rm{C_6H_{12}}$ and $\rm{C_6D_{12}}$ is similar, suggesting that they could play a minor role for the evaluation of isotope effects in this potential energy surface.

\subsection{Analysis of PIMD results} \label{section_analysis_PIMD}

\begin{figure*}[htbp]
    \begin{center}
        \includegraphics[width=0.9\textwidth]{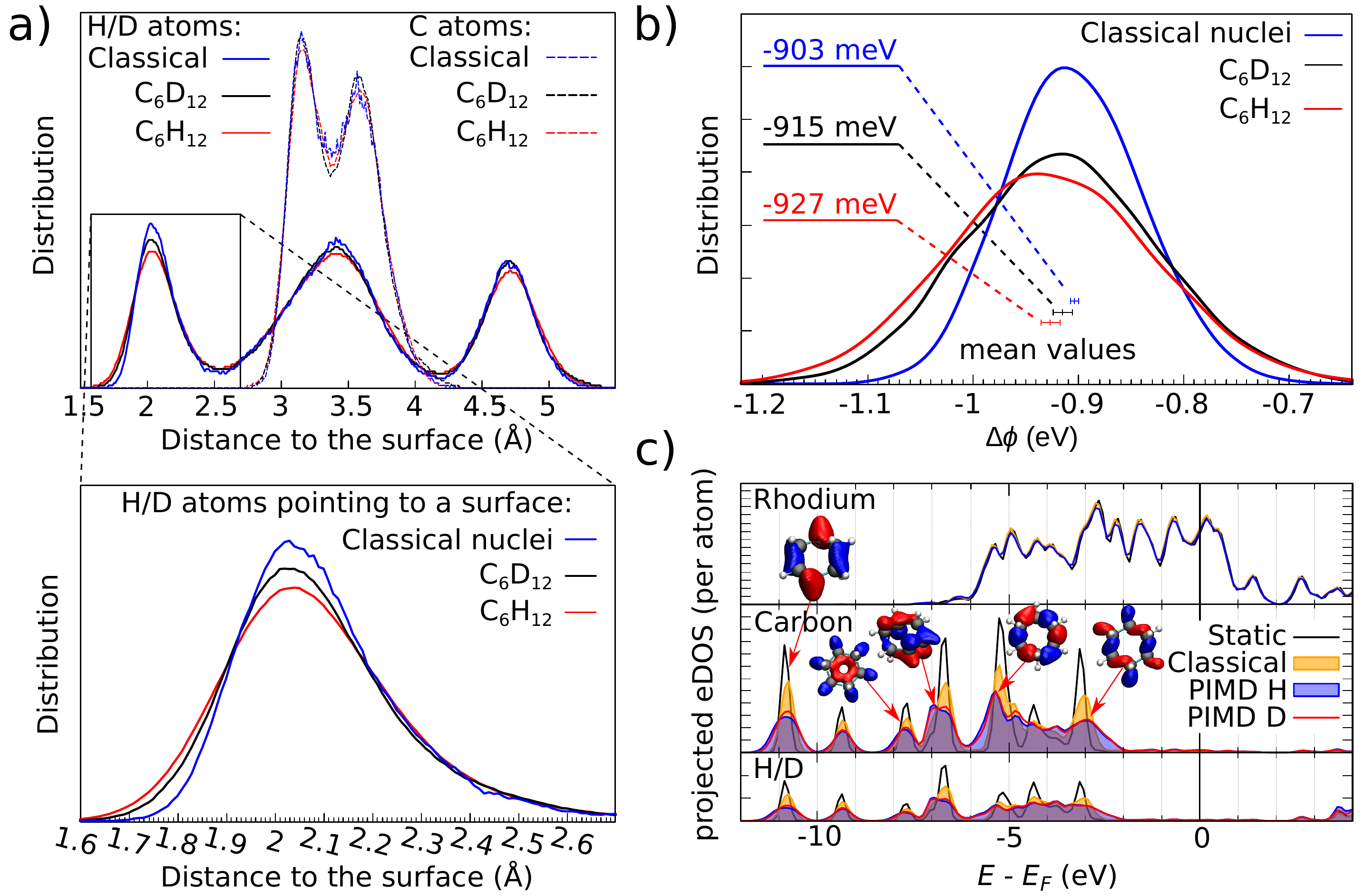}
        \caption{a) The distribution of distances from the Rh(111) surface to H/D atoms (solid lines) and C atoms (dashed lines). The red (black) lines show PIMD simulations of $\rm{C_6H_{12}}$ ($\rm{C_6D_{12}}$), and the blue lines represent MD simulations with classical nuclei.
        b) The distribution 
        of $\Delta \phi$ values for PIMD simulations of $\rm{C_6H_{12}}$ (red), $\rm{C_6D_{12}}$ (black) and classical MD simulation (blue). 
        c) The species-projected electronic density of states in a single-point calculation (black), a classical-nuclei MD simulation (yellow), PIMD simulations for $\rm{C_6H_{12}}$ (blue) and $\rm{C_6D_{12}}$ (red). Peaks are broadened and shifted due to coupling with nuclear vibrations.
        Typical Kohn-Sham eigenstates are shown near the corresponding peaks. In all panels, $T = 150$ K.
        } \label{fig_distributions}
    \end{center}
\end{figure*}

From the aiMD and aiPIMD simulations (PBE+vdW$^{\text{surf}}$) we could directly estimate structural properties of the classical and quantum $\rm{C_6H_{12}}$ and $\rm{C_6D_{12}}$ on Rh(111) at 150 K. In addition, we could capture changes in the electronic structure including full electron-phonon coupling at the adiabatic limit, by averaging the desired electronic quantities of interest over the trajectories. The only drawback, as we will see below, is that even with the SL-RPC technique, statistically converging the quite small energy differences and structural changes observed upon deuteration is a very challenging task. Each force evaluation containing the full interface with the model $\theta = 0.46$ (FHI-aims program, \textit{light} settings) amounts, on average, to 3.1 minutes when parallelized over 240 cores (Intel Xeon Gold 6148 Skylake processors, COBRA supercomputer). This cost renders these simulations computationally expensive even without considering nuclear quantum effects. \rev{Even so, we ensured at least 30 ps of trajectories for each of the systems that we consider (see Methods).}

The results are summarized in figure~\ref{fig_distributions}a-c. In panel a, we show the distribution of the distance from the adsorbate atoms to the top layer of the Rh(111) surface. As expected, a more localized position distribution is observed for $\rm{C_6D_{12}}$ than for $\rm{C_6H_{12}}$, and it is even more localized for classical-nuclei cyclohexane. The inset in panel a shows that $\rm{C_6H_{12}}$ can reach closer to the surface than $\rm{C_6D_{12}}$ and classical-nuclei cyclohexane, but it was not possible to resolve differences on the average position $h_{\rm{COM}}$ to an accuracy of $0.01$ \AA.

In figure~\ref{fig_distributions}b, the distribution of work function values at 150 K is shown.
Again, $\rm{C_6H_{12}}$ presents a broader distribution than $\rm{C_6D_{12}}$ and classical-nuclei MD. 
The distributions are shifted with respect to each other, and their mean values $\langle \Delta \phi \rangle$ are ordered so that $\langle \Delta \phi \rangle_{\rm{H}} < \langle \Delta \phi \rangle_{\rm{D}} < \langle \Delta \phi \rangle_{\rm{Classical}} $. 
The resulting values for $\langle \Delta \phi \rangle$ are \rev{\textbf{--927 $\pm$ 9}~meV for $\rm{C_6H_{12}}$, \textbf{--915 $\pm$ 9}~meV for $\rm{C_6D_{12}}$}, and \textbf{--903 $\pm$ 5}~meV for classical-nuclei cyclohexane. \rev{We were careful in evaluating these uncertainties, by analyzing the autocorrelation behavior of this quantity during the simulation.}
\rev{The H/Classical difference is \textbf{24 $\pm$ 10} meV and we expect the H/D difference to be between zero and this value. In fact, we compute the H/D difference to be  \rev{\textbf{12 $\pm$ 13}} meV, which, despite the large uncertainty, shows the expected trend.} 
Compared to the QH approximation, the aiMD and aiPIMD simulations predict the molecules \rev{(with either classical or quantum nuclei)} to lie farther away from the surface ($h_{\rm{COM}}={\bm{3.42} \pm \bm{0.01}}$~\AA)  by around $\bm{0.06} \pm \bm{0.01}$ \r{A} (See estimation of $h_{\rm{COM}}$ within the QH approximation including temperature effects in \rev{section~4 in the SI}).
Accordingly, the aiPIMD simulations predict a considerably smaller overall work function change. This is a consequence of taking into account anharmonic contributions at a temperature of 150 K (we note that the rigid ``out of plane'' vibrations of the adsorbates lie around 80-130~cm$^{-1}$, thus having components that are thermally activated at 150 K).
This is also consistent with the high anharmonic score of the forces in the $z$ direction, especially at H-bonded sites. Statistically converging the differences between $\rm{C_6H_{12}}$ and $\rm{C_6D_{12}}$ would require a considerable computational effort.
However, with the current uncertainty intervals, it is possible to conclude that the isotope effects from the PIMD simulations cannot differ largely from the QH results, confirming that anharmonic contributions play a minor role \rev{on the geometric isotope effects in} this potential energy surface. \rev{We were not able to calculate the isotope effect on the binding energies because, as mentioned previously, the SL-RPC approximation would have to include (many) more replicas of the system for an accurate assessment, which would make the calculations prohibitive.}

The aiPIMD simulations, \rev{nevertheless}, give access to the electron density of states renormalized by the quantum fluctuations of the molecules. 
\rev{We project the total electronic density of states on the atomic species and average it over multiple snapshots of the simulations. The results are compared with the static density of states in figure~\ref{fig_distributions}c.} There is a pronounced broadening of the peaks only on the adsorbate (for both quantum and classical nuclei), and this broadening is much more pronounced when considering quantum nuclei. We note that it is not clear if one can assign any physical interpretation to such broadening of Kohn-Sham single-particle orbitals. Nevertheless, this effect is caused by the dependence of these ground-state orbital energies on nuclear configurations and the interplay of this dependence with the distribution of nuclear configurations. In addition, there are considerable energy shifts due to this electron-phonon interaction in levels associated with $sp^3$ orbitals. Since $sp^3$ orbitals are responsible for C-H bonding, we tentatively correlate these shifts with the interplay of ZPE and anharmonicity, which effectively changes bond-lengths and thus the electronic structure in this system.
The semilocal/nonlocal functionals we employ are not able to provide a quantitative level alignment of this interface, even if they can predict the HOMO level reasonably well because of the cancellation of the self-interaction error and the missing image-potential effect~\cite{Neaton_2006,Garcia-Lastra_2009}. Even though a much higher level of theory (e.g. many-body perturbation theory) would be desirable for a quantitative comparison with UPS experiments conducted at this interface~\cite{Koitaya_2013}, the magnitude of changes that we observe in the Kohn-Sham electronic density of states highlights the importance of taking nuclear fluctuations into account when analysing the electronic spectra of such interfaces.

\section{Conclusions}

In summary, we have studied isotope and anharmonic effects on the cyclohexane/Rh(111) interface by means of DFT calculations coupled to harmonic lattice dynamics, aiMD, and aiPIMD.

Employing a QH approximation, in which the harmonic ZPE contributions were calculated with the molecule fixed at different distances from the surface, it could be shown that the binding energy of $\rm{C_6D_{12}}$ is smaller than $\rm{C_6H_{12}}$ and that $\rm{C_6D_{12}}$ lies 0.01 \AA~ farther from the surface than $\rm{C_6H_{12}}$, in qualitative agreement with the isotope effects previously observed experimentally~\cite{Koitaya_2012} at the same interface. By showing that the work-function change of the interface is very sensitive to the molecule-surface distance, this geometrical isotope effect could be correlated with the isotope-induced change in the work function, thus confirming the hypothesis that Koitaya, Yoshinobu and coworkers proposed~\cite{Koitaya_2014}, based on experimental observations. 
Finally, these simulations also showed that the electronic-density rearrangement at the interface is impacted by both bond formation and the pushback effect and that the inclusion of van der Waals contributions,
\rev{improve the energetics and adsorption distances.}

The reliability of the QH approximation was assessed by estimating the degree of anharmonicity the nuclear motions at a temperature of 150 K. Anharmonic contributions to the forces are  particularly pronounced at \rev{surface sites that bond to hydrogens} and on the degrees of freedom belonging to the adsorbates. In these cases, in particular, the difference between classical and quantum anharmonic contributions is also large, meaning that techniques like PIMD are necessary to describe structural aspects and related electron-phonon interactions in these systems. However, the \rev{\textit{quantum} part of the} anharmonic contributions to $\rm{C_6H_{12}}$ and $\rm{C_6D_{12}}$ are very similar \rev{in magnitude and character} for coordinates parallel to the surface, and thus play a minor role when addressing isotope effects. This explains why the QH approximation fares well for these quantities \rev{in this case.} 

Indeed, in the \rev{aiMD and} aiPIMD simulations, the pronounced anharmonic character of certain degrees of freedom in the direction perpendicular to the surface plane cause the equilibrium distance of the adsorbates to be around $0.06$ \AA~ farther from the surface than \rev{a static evaluation or the} QH approximation would predict. \rev{This effect stems mostly from anharmonic terms that are already captured with classical nuclei.} This is accompanied by considerably smaller work function changes. However, as expected due to the small contribution of \rev{quantum} anharmonic effects beyond the QH approximation and within the statistical error bars, the observed isotope effects on this system (distance to surface and work function change) do not differ significantly from the QH case. 
Finally, the effect of electron-phonon coupling on the electronic density of states in the adiabatic limit causes a pronounced shift (and broadening) of Kohn-Sham levels related to the CH bonds. 

Although we obtain excellent qualitative agreement with experiment and are able to provide an atomistic view on the origins of the isotope effects measured in this interface, \rev{there are still remaining differences in the magnitude of the isotope effect, in particular on the adsorption energy.}
We are left with the conclusion that this disagreement is likely coming from slightly different conditions in experiment \rev{e.g., clustering of molecules at lower coverages,} or the remaining approximations that were employed in this work. \rev{The most prominent approximations are} the DFT functional and the SL-RPC approximation. \rev{We also cannot rule out that very slow degrees of freedom are not sufficiently sampled within the dynamical simulations.} Nevertheless, we suggest that the \rev{exchange-correlation} functional would be the largest source of remaining errors, given the known drawbacks that functionals based on generalized gradient approximations present for adsorbates on metallic surfaces \cite{MAURER201672}, \rev{and the scatter of binding energies and distances we observed for different functionals and vdW corrections. Moreover, the functionals that yield good binding energies in comparison to experiment, seem to predict a work-function variation with distance to the surface that is too small.}
\rev{All of these observations} motivate the training of fitted or machine-learned potentials \rev{that include long-range electrostatic interactions~\cite{Grisafi:2019ho} and are based on more accurate potential energy surfaces}~\cite{LiuNeaton2017, Hofmann_2013}. Such potentials would both decrease the cost related to statistical sampling and increase the (quantitative) predictive power of these simulations.

\section{Methods}

\textbf{\textit{Electronic structure calculations with FHI-aims:}} Energies and forces are calculated using density-functional theory (DFT) with the PBE exchange-correlation (XC) functional \cite{PBE} and \rev{the range-separated hybrid HSE06 functional \cite{hse06}}. The calculations were done with the all-electron FHI-aims code, which uses numerical atom-centered orbitals~\cite{FHIaims} as basis sets. The FHI-aims package contains predetermined settings for numerical parameters and basis sets, which are aimed at different accuracy levels. 
\textit{Light} settings were used for PIMD and phonon calculations, and \textit{tight} settings were used for potential energy curves and electron density rearrangement. \rev{The parameters of Rh for the pairwise Tkatchenko-Scheffler van der Waals correction modified in order to capture the collective response of a surface in the Lifshitz-Zaremba-Kohn form~\cite{Ruiz2012} (vdW$^{\text{surf}}$) were taken from Ref.~\cite{Ruiz2016}.} The van der Waals interaction between Rh atoms was not included. \rev{Further, we used $\beta=0.81$ for the nl-MBD correction with the PBE functional and $\beta=0.83$ for the same correction with the HSE06 functional.}

We considered 4 Rh(111) layers in all FHI-aims calculations. The 2 bottom layers of the slab are fixed in the bulk geometry. The bulk geometry was calculated for the single-atom FCC unit cell using 16x16x16 k-point grid. \rev{The k-point grid for the different surface unit cells were scaled accordingly. For the $5\times5$ Rh(111) surface unit cell, a $2 \times 2 \times 1$ k-point grid was employed.} The resulting lattice constant of 3.83 \r{A} is in good agreement with the experimental value of 3.80 \r A~\cite{Arblaster_1997}.
The surface was aligned perpendicular to the $z$ axis. In order to isolate the system from its periodic replicas in the $z$ direction, a dipole correction \cite{Neugebauer_Scheffler_1992} and vacuum layer of 30~\r{A} in both directions were applied. 

Vibrational analysis was performed by a modified version of the Phonopy code coupled to  FHI-aims~\cite{phonopy,fidanyan_phonopy}, which allowed to build the Hessian only for the molecular adsorbate and to account for a surface as a rigid environment. 
This approximation is well justified because the coupling between Rh atoms and the molecules is weak, and this weak coupling is concentrated in the low-frequency modes of the adsorbates, which behave very similarly for H- and D-cyclohexane and thus do not impact isotope effects. We set atomic displacements to 0.01 \r{A} for finite difference calculations \rev{and considered geometries relaxed with a maximum force threshold of 0.001 eV/\AA.}. 

\rev{\textbf{\textit{Electronic structure calculations with Quantum Espresso:}}
Calculations were carried out with the rev-vdW-DF2 XC functional~\cite{Hamada_rev_vdW_DF,Callsen_2015}, a variant of the van der Waals density functional\cite{Thonhauser_2007,Berland_2015} as implemented in 
the Quantum ESPRESSO (QE)~\cite{QE_2009, QE_2017} plane-wave code.
QE was used to calculate the adsorption energy and geometry of cyclohexane in a Rh(111) (5$\times$5) surface unit cell ($\theta=0.46$).
The Rh(111) (5$\times$5) surface was modeled by using a 4 atomic layer slab with a vacuum equivalent to 17 monolayers ($\sim$ 40 \r{A}).
The slab was constructed using the lattice constant optimized with rev-vdW-DF2 of 3.80 \r{A}.
The molecule was put on one side of the slab and the effective screening medium method was used to eliminate the artificial electrostatic interaction with the neighboring slabs
~\cite{Otani_Sugino_2006, Hamada_screening_2009}.
The molecular geometry was optimized starting from that determined by FHI-aims.
The bottom 2 layers of the slab were fixed and the remaining degrees of freedom were optimized until the force acting on them were less than $5 \times 10^{-4}$ Ry/Bohr (1.3 $\times 10^{-2}$eV/\r{A}).
Projector augmented wave~\cite{Bloechl_1994} potentials from the pslibrary version 1.0.0~\cite{Dal_Corso_2014} and a plane-wave basis set with a cutoff energy of 80 (640) Ry were used for wave functions (augmentation charge density).
A $2 \times 2$ k-point grid was used to sample the surface Brillouin zone.}

\textit{\textbf{Adsorption energies and free energies:}} 
The adsorption energies per molecule were calculated with
\begin{equation}
    E_{\text{ads}}^{\text{pot}} = (E^{\text{pot}}_{\text{s+m}} - E^{\text{pot}}_{\text{s}})/N_{\rm{mol}} - E^{\text{pot}}_{\text{m}} \label{eq_Eads},
\end{equation}
where $E^{\text{pot}}_{\text{s+m}}$ is the total energy at the potential energy surface of a slab with molecules adsorbed,
$E^{\text{pot}}_{\text{s}}$ is the total energy of a clean surface relaxed with 2 bottom layers fixed in bulk position,
$E^{\text{pot}}_{\text{m}}$ is the total energy of a molecule relaxed in vacuum, 
and $N_{\rm{mol}}$ is the number of molecules in a unit cell.
A similar expression can be written for a free energy of adsorption \rev{$F^{\rm{harm}}_{\rm{ads}}$}.

The harmonic vibrational free energy was calculated as
\begin{equation}
\begin{split}
    \rev{F^{\rm{harm}}} = 
    & \sum_{i=1}^{N_{\rm{modes}}} \left[ \frac{\hbar \omega_i} {2} + k_B T \ln \left (1-\exp^{-\frac{\hbar \omega_i}{k_B T}} \right) \right ] + \\
    & +_{\text{[if gas phase]}} \left(F^{\rm{trans}} + F^{\rm{rot}}\right), \label{eq_Fharm}
\end{split}
\end{equation} 
where $N_{\rm{modes}}=3N - 3$ when the free energy of a clean surface is calculated ($N$ is the number of atoms in a unit cell), $N_{\rm{modes}}=3N - 3N_{\rm{s}}$ ($N_s$ is the number of surface atoms in a unit cell) when the free energy of molecules adsorbed on surface is calculated, and $3N_m - 6$ ($N_m$ is the number of atoms in a molecule), when the free energy of an isolated molecule is calculated. Rotational and translational contributions were added for the free molecule according to the rigid-body and ideal gas textbook expressions~\cite{Landau_Lifshitz_vol5}. \rev{The expression above only takes into account vibrations at the $\Gamma$ point of the unit cell. Because we focus mostly on molecular vibrations, \rev{which show a very small phonon band dispersion, and employ large unit cells}, this approximation does not introduce large errors in the calculated free energy differences.}
For the translational term, we took a pressure of $10^{-8}$ Pa, which is close to the reported experimental conditions~\cite{Koitaya_2012}.

The quasi-harmonic ZPE-corrected energy of adsorption $E^{*}_{\rm{ads}}$ was calculated as a difference between the energy at the equilibrium distance and at $10$~\r{A} away from the surface, which is considered to be far enough to remove all molecule-surface interaction,
\begin{equation} 
    \begin{split}
        E^{*}_{\rm{ads}}(h_{\rm{COM}}) =
      & \left(E^{\rm{pot}}_{\text{s+m}}(h_{\rm{COM}})
        + \sum_{i=1}^{3N_m -3} \frac{\hbar \omega_i} {2}(h_{\rm{COM}})\right) - \\
      & - \left.\left( E^{\rm{pot}}_{\text{s+m}}
        + \sum_{i=1}^{3N_m -3} \frac{\hbar \omega_i} {2}\right) 
        \right\vert_{h_{\rm{COM}} = 10\text{\AA}}, \label{eq:quasiharm-eads}
    \end{split}
\end{equation}
where $h_{\rm{COM}}$ denotes the  distance from the center of mass of the adsorbate to the Rh surface. Although $E^{*}_{\rm{ads}}$ is not a true adsorption energy, this procedure compensates spurious interactions that might appear in a particular simulation cell.
For the ZPE contribution, we include $(3N_m -3)$ molecular vibrations.

\textbf{\textit{Ab initio molecular dynamics:}} AiMD and aiPIMD simulations were carried out by connecting FHI-aims to the i-PI code~\cite{iPIv2}. For classical-nuclei MD, a timestep of 1 fs was used. For PIMD simulations, a smaller timestep of 0.5 fs was employed.

In order to accelerate sampling in the NVT ensemble, we applied a colored-noise generalized Langevin thermostat (GLE)\cite{GLE2009, GLE2010} to the classical-nuclei aiMD simulations.
For the aiPIMD simulations, the PIGLET thermostat was used~\cite{PIGLET}.
This approach preserves quantum distribution and gives a fast convergence of observables with respect to the number of replicas.
The parameters for the thermostats are: 8 fictitious degrees of freedom $s$ ~\cite{GLE2010}, and a frequency range of 0.32-3200 cm$^{-1}$ for $\rm{C_6H_{12}}$ and 0.23-2300 cm$^{-1}$ for $\rm{C_6D_{12}}$. The \textbf{A} and \textbf{C} matrices (as defined in~\cite{GLE2010}) are parameterized for $\hbar \omega / k_B T = 50$ using the GLE4MD library~\cite{gle4md}. We observed convergence with around 12 beads for the adsorbate atoms (H) at 150 K within this setup.
\rev{After thermalization,} we have calculated \rev{7} independent trajectories of total length of \rev{32} ps for $\rm{C_6H_{12}}$, 5 trajectories of total length of \rev{30} ps for $\rm{C_6D_{12}}$, and 5 trajectories of total length of 97 ps for classical-nuclei cyclohexane. 

The effects of nuclear fluctuations on electronic observables (work function, electronic density of states, etc.) were calculated as the average of single-point calculations from aiPIMD trajectories through the following general expression
\begin{equation} \label{eq_observables}
\begin{split}
     \langle A \rangle = 
    & \frac{1}{Z} \mathrm{Tr} \left[ \hat{A} e^{\frac{-\hat{H}}{k_B T}} \right] 
     \xRightarrow[\text{PIMD sampling}]{\text{ergodicity,}}  \\
    & \xRightarrow{} \frac{1}{P N_s} \sum_{i}^{N_t} \sum_{k}^{P} A \left(\bm{q}^{(k)}(t_i)\right),
\end{split}
\end{equation}
where $\hat{A}$ is a position-dependent observable, $\hat{H}$ is the Hamiltonian,
$Z$ is the partition function,
$P$ is the number of beads of a ring polymer,
$N_t$ is the number of snapshots considered from a PIMD trajectory, and
$\bm{q}^{(k)}$ is a position vector for a bead $k$.
The snapshots from the PIMD trajectory were picked so that they were statistically independent. The criterion for independence is an autocorrelation time of the property A. In our PIMD calculations, the autocorrelation time is 30 fs for a velocity, 300 fs for a work function and 600 fs for the z-coordinate of the center of mass of a molecule. We also used these correlation times to calculate error bars. For classical nuclei simulations, the same expression was used with $P=1$.

\section{Acknowledgements}
We acknowledge fruitful and enlightening discussions with Prof. Jun Yoshinobu and Prof. Takanori Koitaya, who have also motivated this work with their experiments and initial hypothesis. M.R. and K.F. further acknowledge lively scientific discussions about this work with Dmitrii Maksimov and Yair Litman. I.H. acknowledges discussions with Prof. Toshiki Sugimoto. Most of the computational time was provided by the Max Planck Computing and Data Facility (MPCDF), and part of the computation was performed at Research Institute for Information Technology, Kyushu University.

M.R. and K.F. acknowledge financial support from the International Max Planck Research School on Functionalized Interfaces. I.H. acknowledges financial support from Japan Society for the Promotion of Science through Grant-in-Aid for Scientific Research on Innovative Areas "Hydrogenomics" (Grant No. JP18H05519).

\bibliography{references}

\end{document}

\title{Supplemental information \\  Quantum Nuclei at Weakly Bonded Interfaces: The Case of Cyclohexane on Rh(111)}

\author{Karen Fidanyan}
\affiliation{Fritz Haber Institute of the Max Planck Society, Faradayweg 4-6, 14195 Berlin, Germany}
\affiliation{Max Planck Institute for the Structure and Dynamics of Matter, Luruper Chaussee 149, 22761 Hamburg, Germany}

\author{Ikutaro Hamada}
\email{ihamada@prec.eng.osaka-u.ac.jp}
\affiliation{Department of Precision Engineering, Graduate School of Engineering, Osaka University, 2-1 Yamadaoka, Suita, Osaka 565-0871, Japan}

\author{Mariana Rossi}
\email{mariana.rossi@mpsd.mpg.de}
\affiliation{Fritz Haber Institute of the Max Planck Society, Faradayweg 4-6, 14195 Berlin, Germany}
\affiliation{Max Planck Institute for the Structure and Dynamics of Matter, Luruper Chaussee 149, 22761 Hamburg, Germany}

\maketitle

\section{Adsorption properties}
Adsorption energies $E_{\text{ads}}^{\text{pot}} $ for the systems discussed in the paper are provided in table~\ref{table_Eads_coverage}.
\begin{table}[ht]
\caption{Adsorption energy for different adsorption patterns, calculated with PBE+vdW$^{\rm{surf}}$ by FHI-aims code with \textit{Light} and \textit{Tight} settings.} \label{table_Eads_coverage}
\begin{center}
\begin{tabular}{|l|r|}
\hline
 & $\bm{E_{\rm{ads}}}$ $\left(\rm\frac{\textbf{eV}}{\textbf{molecule}} \right)$ \\ \hline
$\bm {\theta = 1, \left(2\sqrt{3}\times 2\sqrt{3}\right)R13.9\degree,}$ \textbf{9 molecules, 4 Rh layers} &  \\
\textit{Light}, 2x2x1 k-points & 1.023 \\ \hline
%\textit{Tight}, only $\Gamma$-point & 1.002 \\
%\textit{Tight}, 3x3x1 k-points &  0.979 \\ \hline
% $\bm{\theta = 0.84, (2.64\times2.64)R19.1 \degree},$ \textbf{7 molecules, 4 Rh layers} &  \\
% \textit{Tight}, only $\Gamma$-point & 1.020 \\
% \textit{Tight}, 3x3x1 k-points & 0.976 \\ \hline
$\bm{\theta = 0.64},$ \textbf{3x3 slab with 1 molecule} & \\
\textit{Light}, 4x4x1 k-points &  0.946 \\ \hline
$\bm{\theta = 0.46},$ \textbf{5x5 slab with 2 molecules} &  \\
\textit{Light}, 2x2x1 k-points & 0.953 \\
\textit{Tight}, 2x2x1 k-points & 0.912 \\ \hline
$\bm{\theta = 0.12},$ \textbf{7x7 slab with 1 molecule} & \\
\textit{Light}, 2x2x1 k-points &  0.945 \\ \hline
\end{tabular}
\end{center}
\end{table}

\begin{figure}
    \centering
    \includegraphics[width=0.6\textwidth]{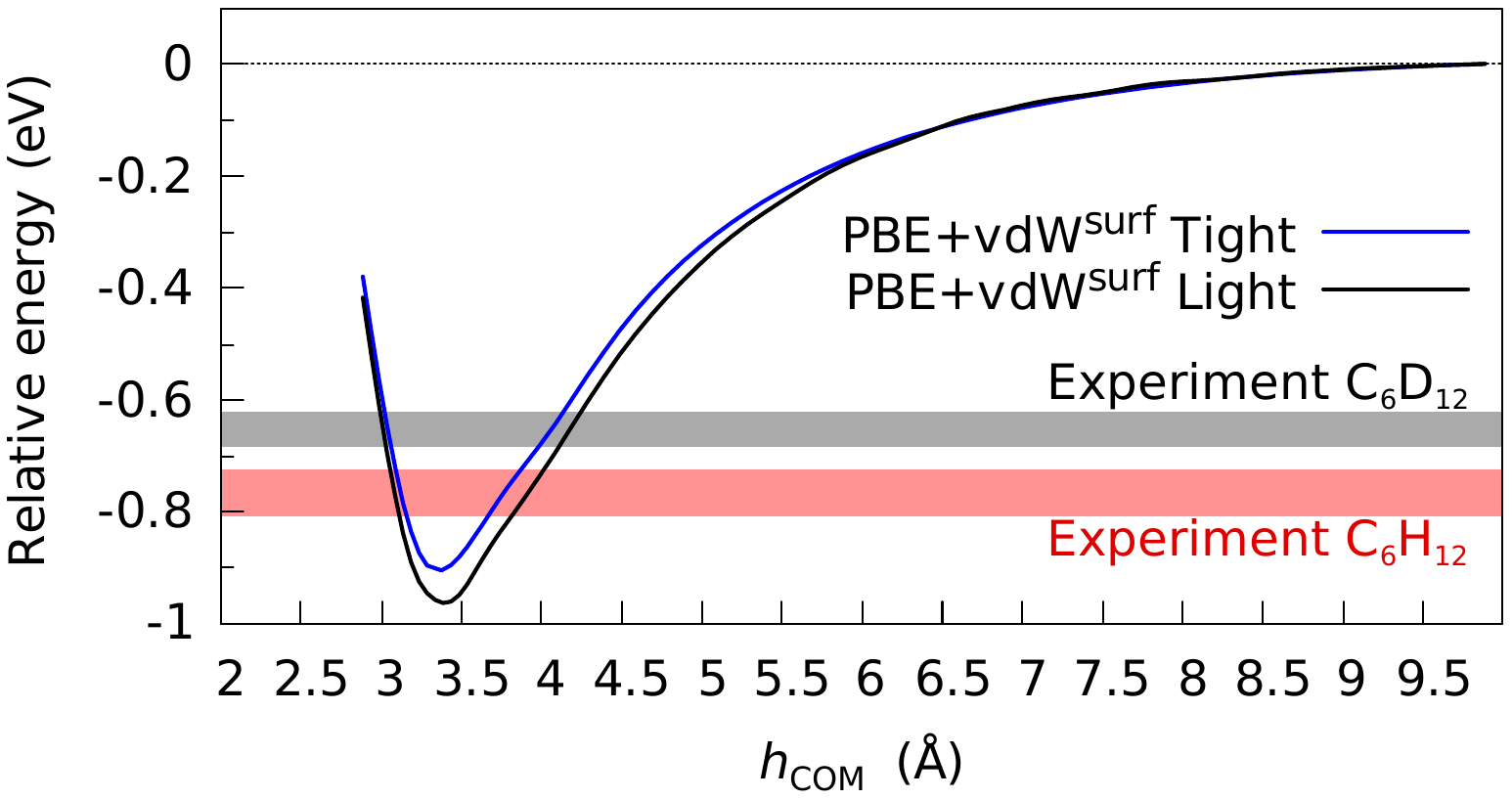}
    \caption{Adsorption curves calculated with PBE +$\rm{vdW^{surf}}$~\cite{Ruiz2012} functional with \textit{Tight} (solid blue line) and \textit{Light} (dashed black line) settings of FHI-aims. Calculations were performed with the unit cell of  $\theta = 0.46$.
    Shaded areas show the experimental values of the adsorption energy of $\rm{C_6H_{12}}$ (red) and  $\rm{C_6D_{12}}$ (grey), obtained by temperature programmed desorption~\cite{Koitaya_2012}.}
    \label{fig_E_z_light-tight}
\end{figure}

A comparison of the adsorption energy curve between \textit{Light} and \textit{Tight} settings is provided in figure~\ref{fig_E_z_light-tight} for $\theta=0.46$. The difference in binding energy is 58 meV.

\rev{As mentioned in the paper, we use $E_{\rm{ads}}$ calculated by the eq. 5 and $E^*_{ads}$ calculated by the eq. 7. Only $E_{\rm{ads}}$ rigorously satisfies the definition of an adsorption energy.
Comparing the numbers for coverage 0.46 and $\rm{PBE+vdW^{surf}}$ in  tables~\ref{table_Eads_coverage} and \ref{tab_Eads_and_dist_all_xc}, one can compare the values obtained with these two definitions. The differences are very small (less than 10 meV).}

%\rev{Another detail to mention is that the adsorption curves presented in the paper were calculated with the surface atoms fixed at equilibrium positions. 
%In the table~\ref{tab_Eads_and_dist_all_xc} we provide a comparison of fixed/relaxed surface for the coverage of 0.46 and $\rm{PBE+vdW^{surf}}$, \textit{Light} settings. The difference in $E^*_{\rm{ads}}$ is \textbf{7} meV.}
\begin{table}[ht]
    \centering
    \caption{Adsorption distance $h_{\rm{COM}}$ between the center of mass of a molecule and a surface, and energy $E^{*}_{\rm{ads}}$, calculated as the energy difference between the minimal point of the adsorption curve and the point at 10 \r{A} distance from the surface.}
    \label{tab_Eads_and_dist_all_xc}
    \begin{tabular}{l|r|r}
    \textbf{FHI-aims} & $h_{\rm{COM}}$ (\r{A}) & $E^{*}_{\rm{ads}}$ (meV) \\ \hline
    \multicolumn{3}{l}{\textbf{$\theta = 0.46$, \textit{Tight} settings}} \\ 
    PBE           & 3.71  & 147  \\
    PBE+MBD-nl    & 3.45  & 843  \\
    $\rm{PBE+vdW^{surf}}$   & 3.36  & 905  \\
    HSE06+MBD-nl  & 3.30  & 1026  \\ 
    \hline
    \multicolumn{3}{l}{\textbf{$\theta = 0.46$, \textit{Light} settings}} \\
    $\rm{PBE+vdW^{surf}}$   & 3.37  & 963   \\
    %\multicolumn{3}{l}{\textbf{$\theta = 0.46$, \textit{Light}, Rh surface is relaxed at each point}} \\
    %$\rm{PBE+vdW^{surf}}$   & 3.38  & 956   \\
    \hline
    \end{tabular}
\end{table}

\section{Coverage dependence of vibrational spectrum}
\begin{figure}[ht]
    \begin{center}
        \includegraphics[width=0.6\textwidth]{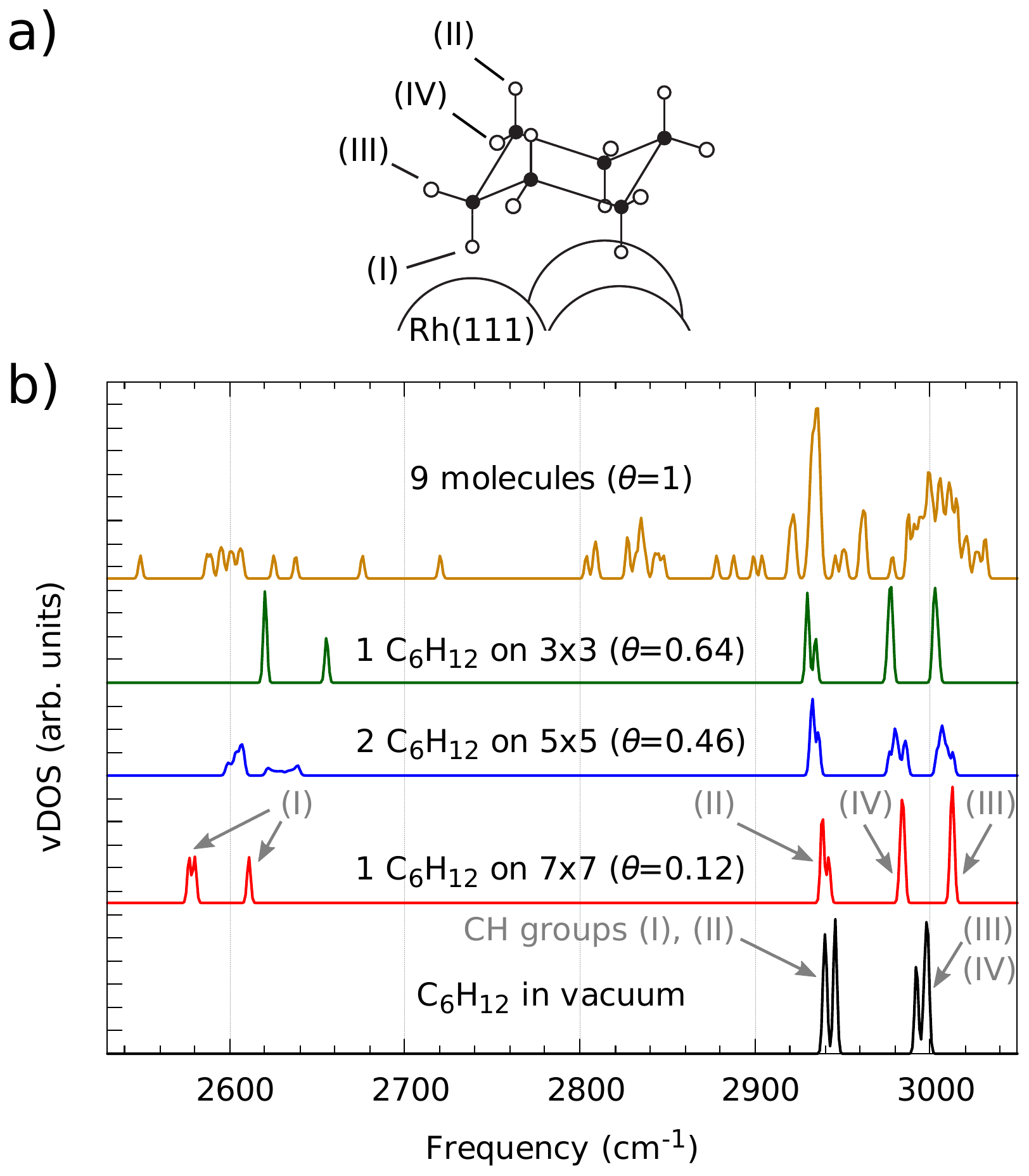}
    \end{center}
    \caption{ 
    a) Different CH groups for a cyclohexane adsorbed on a surface.
    b) The vibrational spectra of CH stretching modes of cyclohexane in vacuum (black) and on a Rh(111) surface with coverage $\theta = 0.12$ (red), $\theta = 0.46$ (blue), $\theta = 0.64$ (green) and $\theta = 1$ (ochre). 
    The grey arrows assign peaks to the CH groups given in (b).
    As the red shift in CH stretching modes decreases, the H/D difference in the adsorption energy decreases also. At the full coverage ($\theta=1$), the intermolecular interaction is so strong that single adsorption sites become highly non-equivalent, which is reflected in multiple peak splitting in the range between 2540 and 3040 cm$^{-1}$.
    } \label{fig_vdos_coverage_dep}
\end{figure}

 We show in figure~\ref{fig_vdos_coverage_dep} the harmonic vDOS for the adsorption patterns considered in this work (PBE+vdW$^{\text{surf}}$). The key feature is a red shift in the stretching vibrations of the CH groups pointing to the surface (denoted as (I) in figure~\ref{fig_vdos_coverage_dep}b). This red shift decreases with increasing coverage. It means that surface-molecule interaction becomes weaker. At highest coverage ($\theta = 1$) there are multiple splittings of CH stretching frequencies. It is an evidence that cyclohexane adsorption sites become highly nonequivalent because of molecule-molecule interactions.
We also note a blue shift in stretching vibrations of CH groups of type (III) and slight red shift in groups (II) and (IV).

\section{Calculations with rev-vdW-DF2 functional}

\begin{figure}[ht]
    \centering
    \includegraphics[width=0.6\textwidth]{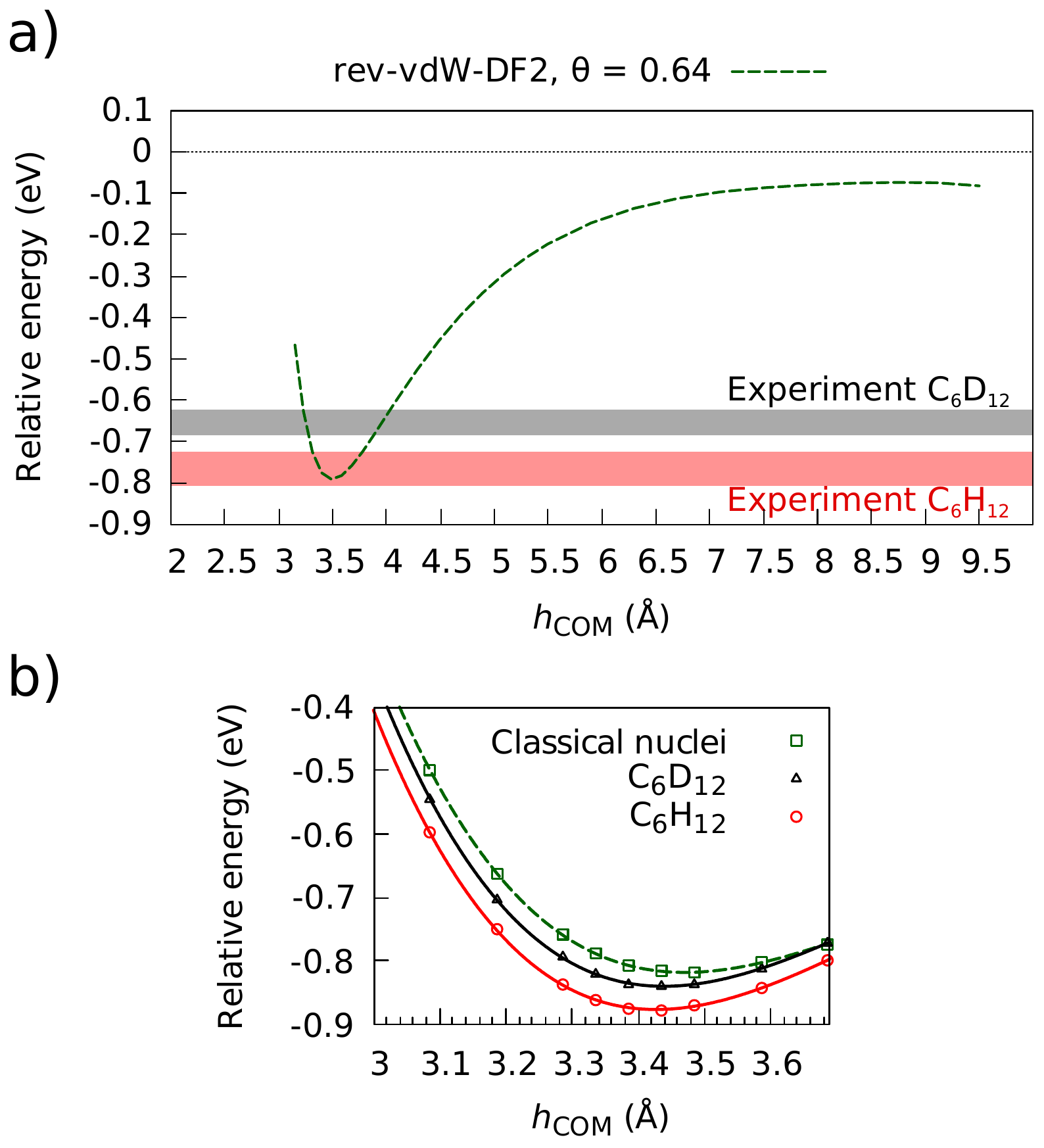}
    \caption{\rev{
    a) Adsorption curve calculated with rev-vdW-DF2 for the coverage of $\theta = 0.64$ obtained using Quantum-ESPRESSO. Zero energy level corresponds to sum of the energies of an isolated molecule and a clean surface.
    Shaded areas show the experimental values of the adsorption energy of $\rm{C_6H_{12}}$ (red) and  $\rm{C_6D_{12}}$ (grey), obtained by temperature programmed desorption~\cite{Koitaya_2012}.
    b) ZPE-corrected energy of adsorption for $\rm{C_6H_{12}}$ (red) and $\rm{C_6D_{12}}$ (black) obtained with STATE. The green line shows the adsorption energy values calculated without ZPE correction. Zero level is the same as in a).}}
    \label{fig_QH_Ikutaro}
\end{figure}

\rev{The Quantum-ESPRESSO code~\cite{QE_2009,QE_2017} was used for the calculations of adsorption energy curve without zero-point-energy correction as a function of molecule-surface distance with a Rh(111) (3$\times$3) surface with 4 metal layers and vacuum equivalent to eight monolayer (19.77 \r{A}) with rev-vdW-DF2 functional as shown in figure~\ref{fig_QH_Ikutaro} (a) ($\theta = 0.64$). A $6\times 6$ $\Gamma$-centered k-point grid was used. The results for $\theta =0.46$ and $\theta = 0.64$ are given in table~\ref{tab_Eads_QE}.}

\rev{The STATE~\cite{Morikawa_2004,STATE} code was used for the calculations of vibrational spectra, adsorption energy curves with and without zero-point-energy correction, and the work function as functions of the molecule-surface distance with the rev-vdW-DF2 functional\cite{Hamada_rev_vdW_DF,Callsen_2015}.
A simplified version of the self-consistent van der Waals density functional\cite{Roman-Perez_2009,Wu_2012} was employed\cite{Hamamoto_2016}.}
\rev{The computational setup for the STATE calculations has multiple differences from the one reported in the main text. For these calculations, a Rh(111) (3$\times$3) surface with 5 metal layers with a vacuum equivalent to eight monolayer (19.88 \r{A}) is considered ($\theta =0.64$).
The slab was constructed using the lattice constant optimized with rev-vdW-DF2 of 3.81 \r{A}.
A cyclohexane molecule was put on one side of the slab and the effective screening medium method was used to eliminate the artificial electrostatic interaction with the neighboring slabs
~\cite{Otani_Sugino_2006, Hamada_screening_2009}. 
The bottom 2 layers of the Rh(111) slab were fixed and the remaining degrees of freedom were fully relaxed until the forces acting on them become less than $5\times 10^{-4}$ Hartree/Bohr.
Ultrasoft pseudopotentials~\cite{Vanderbilt_1990} and a plane-wave basis sets with a cutoff energy of 49 (625) Ry for wave functions (charge density) were used.
A $4 \times 4$ $\Gamma$-centered k-point grid was used.
Harmonic vibrational frequencies were calculated by using the Wilson's GF method and finite difference method as implemented in the STATE package\cite{Wilson_1980}.
Atoms were displaced along the vibrational eigenmodes and the magnitude of the displacement was determined self-consistently. All $3N$ molecular vibrations are included in zero-point energy calculation.}
\rev{The results are presented in  figure~\ref{fig_QH_Ikutaro} (b) and in table~\ref{table_qh_064}.}

% TABLE E_ads Quantum ESPRESSO
\begin{table}[ht]
    \centering
    \caption{Adsorption distance $h_{\rm{COM}}$ between the center of mass of a molecule and a surface, and energy $E_{\rm{ads}}$, calculated as the energy difference between the minimal point of the adsorption curve and the sum of energies of an isolated molecule and a clean surface using Quantum ESPRESSO.}
    \label{tab_Eads_QE}
    \begin{tabular}{l|r r}
    \hline
    \textbf{Quantum ESPRESSO} & $h_{\rm{COM}}$ (\r{A}) & $E_{\rm{ads}}$ (meV) \\ \hline
    $\theta = 0.46$, rev-vdW-DF2 & 3.48  & 768  \\
    $\theta = 0.64$, rev-vdW-DF2 & 3.49  & 791  \\
    \hline
    \end{tabular}
\end{table}

% TABLE QH 0.64
\begin{table}[ht]
\begin{center}
\caption{The results of quasi-harmonic calculation with rev-vdW-DF2: the adsorption energy, the equilibrium distance between molecules and the surface, and the work function change for $\rm{C_6H_{12}}$ and $\rm{C_6D_{12}}$ for coverage $\theta = 0.64$ using STATE.} 
\label{table_qh_064}
\begin{tabular}{r|r r r}
\hline
 \textbf{STATE} & $\rm{C_6H_{12}}$ & $\rm{C_6D_{12}}$ & no ZPE \\ \hline
ZPE-corrected $E_{\rm{ads}}$ (meV) & \rev{876} & \rev{840} & \rev{818} \\
QH $h_{\rm{COM}}$ (\r{A}) & 3.427 & 3.439 & 3.471 \\
QH $\Delta \phi$ (meV)    & -1205 & -1190 & -1151 \\
\hline
\end{tabular}
\end{center}
\end{table}

% TABLE QH + PIMD 0.46
\begin{table}[ht]
\begin{center}
\caption{Isotope effects on distance to surface and work function change, obtained by QH model and aiPIMD simulations with $\rm{PBE + vdW^{surf}}$ functional for coverage $\theta = 0.46$.} 
\label{table_qh}
\begin{tabular}{r|r r r}
\hline
 & $\rm{C_6H_{12}}$ & $\rm{C_6D_{12}}$ & no ZPE \\ \hline
ZPE-corrected $E^{*}_{\rm{ads}}$ (meV) & \rev{1067} & \rev{1030} & \rev{956} \\
QH $h_{\rm{COM}}$ (\r{A}) & \rev{3.351} & \rev{3.364} & \rev{3.386} \\
QH $\Delta \phi$ (meV) & \rev{-960} & \rev{-945} & \rev{-920} \\
\hline
PIMD $h_{\rm{COM}}$ (\r{A}) & 3.41 $\pm$ 0.01 & 3.42 $\pm$ 0.01 & 3.416 $\pm$ 0.007 \\
PIMD $\Delta \phi$ (meV) & \rev{-927 $\pm$ 9} & \rev{-915 $\pm$ 9} & -903 $\pm$ 5 \\
\hline
\end{tabular}
\end{center}
\end{table}

\section{Quasi-harmonic model at finite temperature}
For the calculation of a ZPE corrected adsorption curves, with FHI-aims we include only $(3N_m -3)$ molecular vibrations, where $N_m$ is the number of atoms in a molecule. The remaining 3 lowest-frequency modes (0 to 55 cm$^{-1}$ depending on distance to surface) correspond to the hindered translations of a rigid molecule. They have completely classical behavior and high entropy, and they are populated at very low temperatures. 
These modes give a relatively high mobility to molecules. They raise the question of applicability of the harmonic approximation when the molecule deviates far away from its equilibrium position. 

\begin{figure}[ht]
    \centering
    \includegraphics[width=0.8\textwidth]{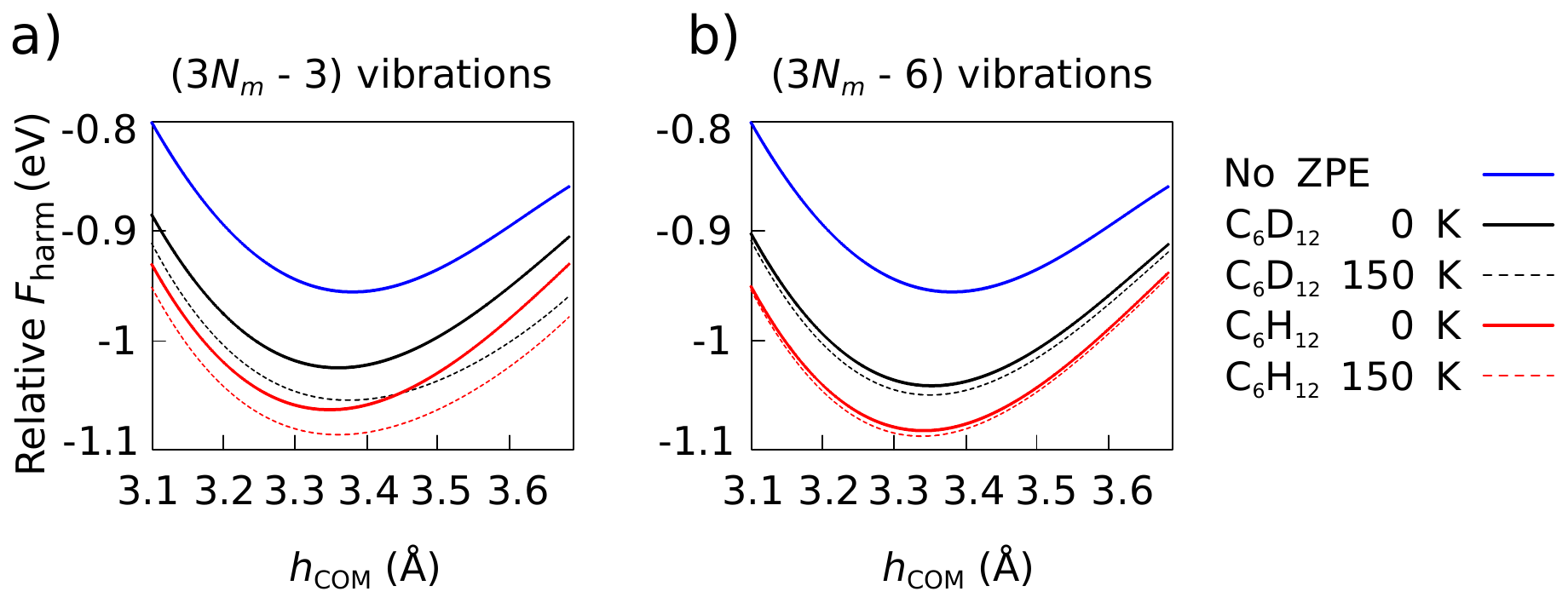}
    \caption{The effect of temperature on the harmonic free energy of cyclohexane (red) and D-cyclohexane (black) with and without inclusion of hindered rigid rotation modes (a and b, respectively).
    Solid lines show ZPE-corrected potential energy, dashed lines add finite temperature corrections at the temperature of 150~K. The curves are aligned \rev{to zero} at the distance of 10~\r{A}. Calculations are done with PBE + $\rm{vdW^{surf}}$ functional.
    } \label{fig_SI_harm-temp}
\end{figure}

In figures~\ref{fig_SI_harm-temp}a,b, we show harmonic free energy corrections at 150~K including $(3N_m -3)$ modes and only the $(3N_m -6)$ intramolecular modes. Clearly, temperature effects are concentrated in the lower-frequency modes, and these effects are negligibly small in the higher frequency intramolecular vibrations. $h_{\rm{COM}}$ when including $(3N_m-3)$ modes is \rev{\textbf{3.351 (3.365)}~\r{A} for H and \textbf{3.364 (3.384)}~\r{A} for D at the temperature of 0 (150) K.}

\section{Dependence of the work function change on coverage and functional}
\begin{figure}[ht]
    \centering
    \includegraphics[width=0.6\textwidth]{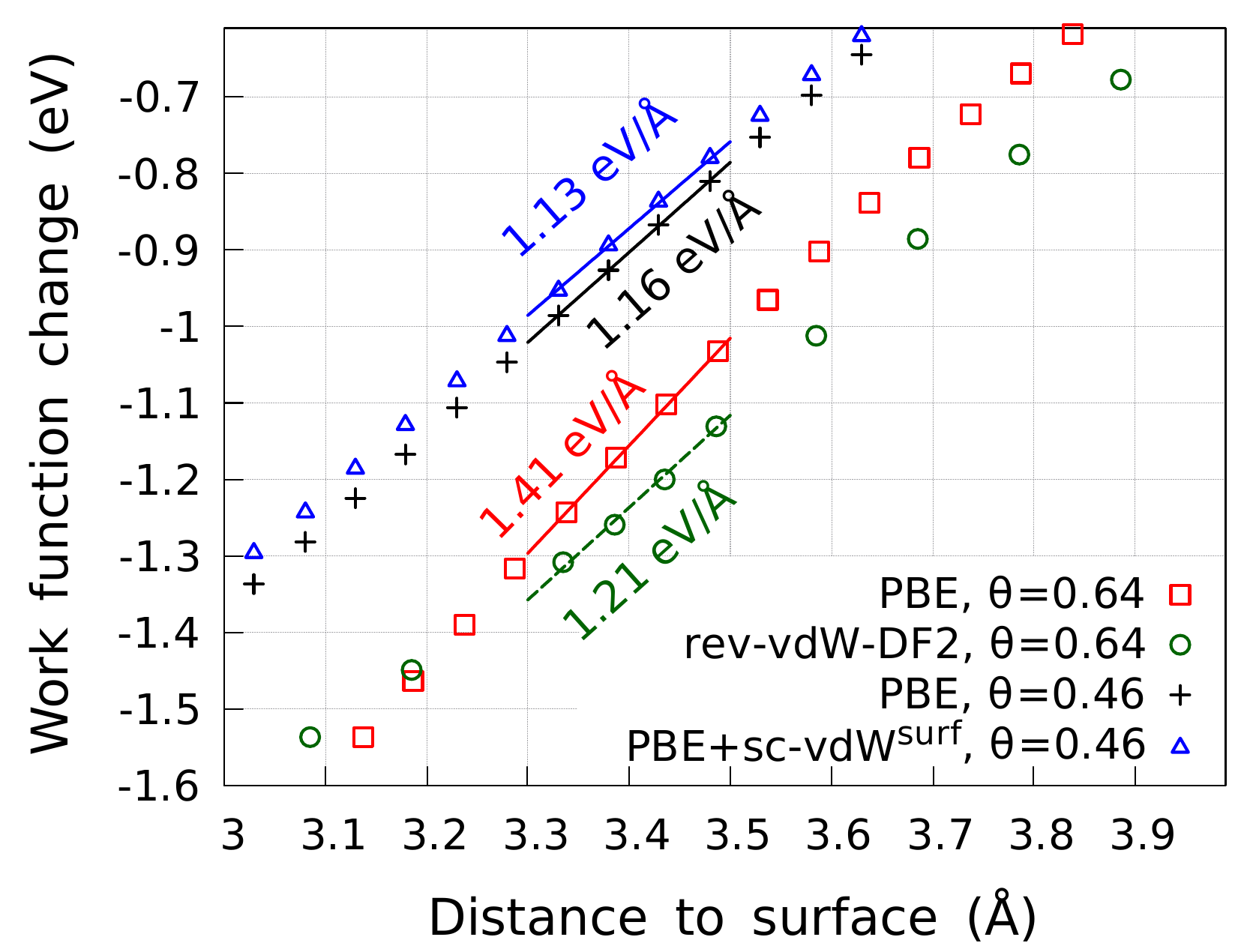}
    \caption{\rev{Work function change as a function of distance to surface. Calculated for coverage 0.64 using PBE (red squares) and rev-vdW-DF2 (green circles) functionals. And the values for coverage 0.46 by PBE (black crosses) and PBE + self-consistent $\rm{vdW^{surf}}$ (blue triangles).}
    } \label{fig_wf_z_3x3}
\end{figure}

\rev{We have calculated the dependence of work function change on distance for the coverage 0.64 using rev-vdW-DF2 and PBE functionals. The results are shown in figure~\ref{fig_wf_z_3x3}. Taking the slope of the rev-vdW-DF2 dependence, the QH model predicts the isotopic difference in work function change to be \textbf{15 meV} (see table~\ref{table_qh_064}).
In addition, we show the difference between self-consistent and non-self-consistent implementations of $\rm{vdW^{surf}}$ correction in the work function~\cite{scvdW}.}

\section{Electron density rearrangement}
\begin{figure}[ht]
    \centering
    \includegraphics[width=0.8\textwidth]{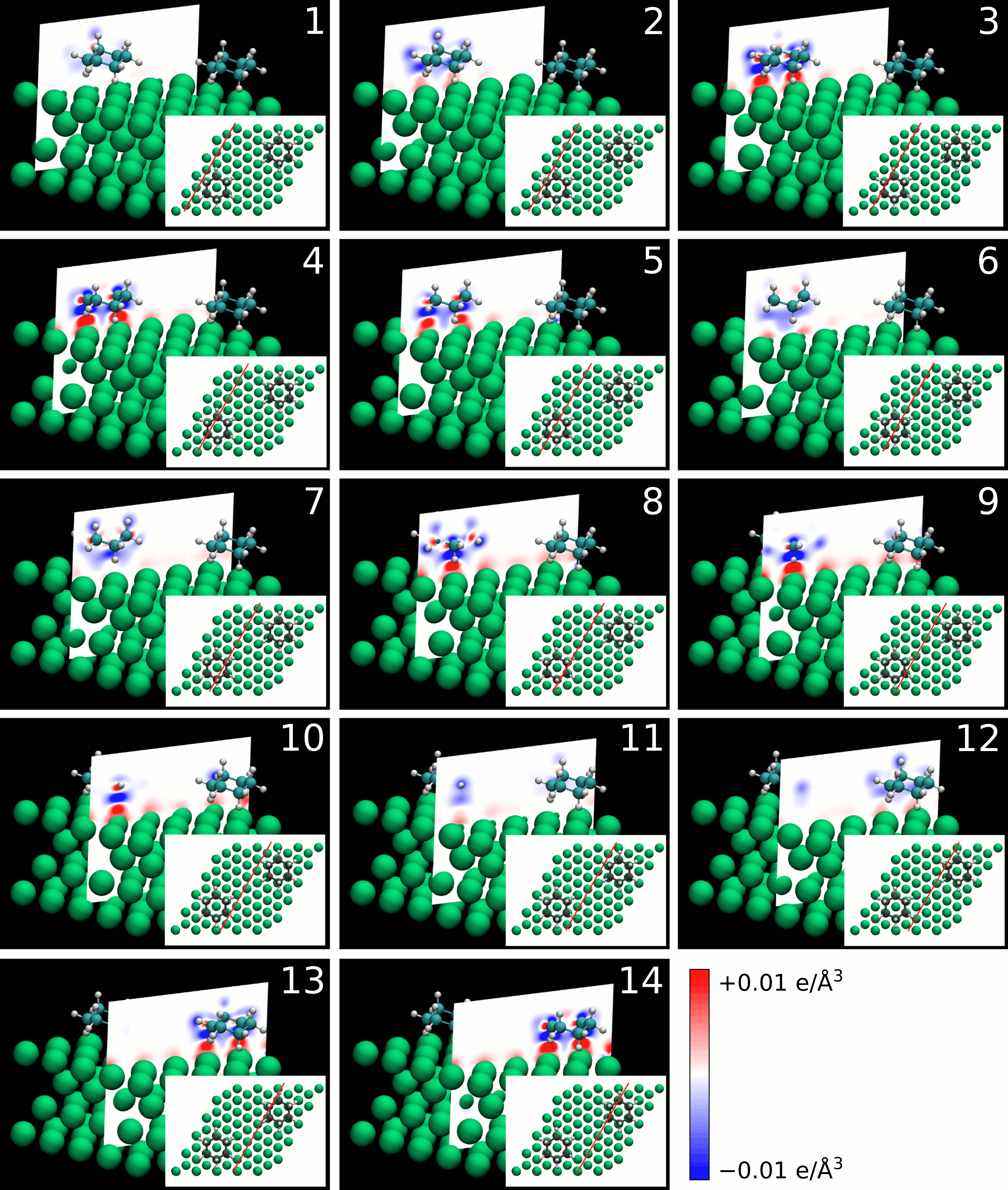}
    \caption{Difference between the electron density of a surface with molecules adsorbed and the sum of isolated surface and isolated molecules, shown at different $y-z$ slices along $x$ coordinate. Red color denotes electron density accumulation, and blue denotes depletion.}
    \label{fig_SI_slices_x}
\end{figure}

\begin{figure}
    \centering
    \includegraphics[width=0.7\textwidth]{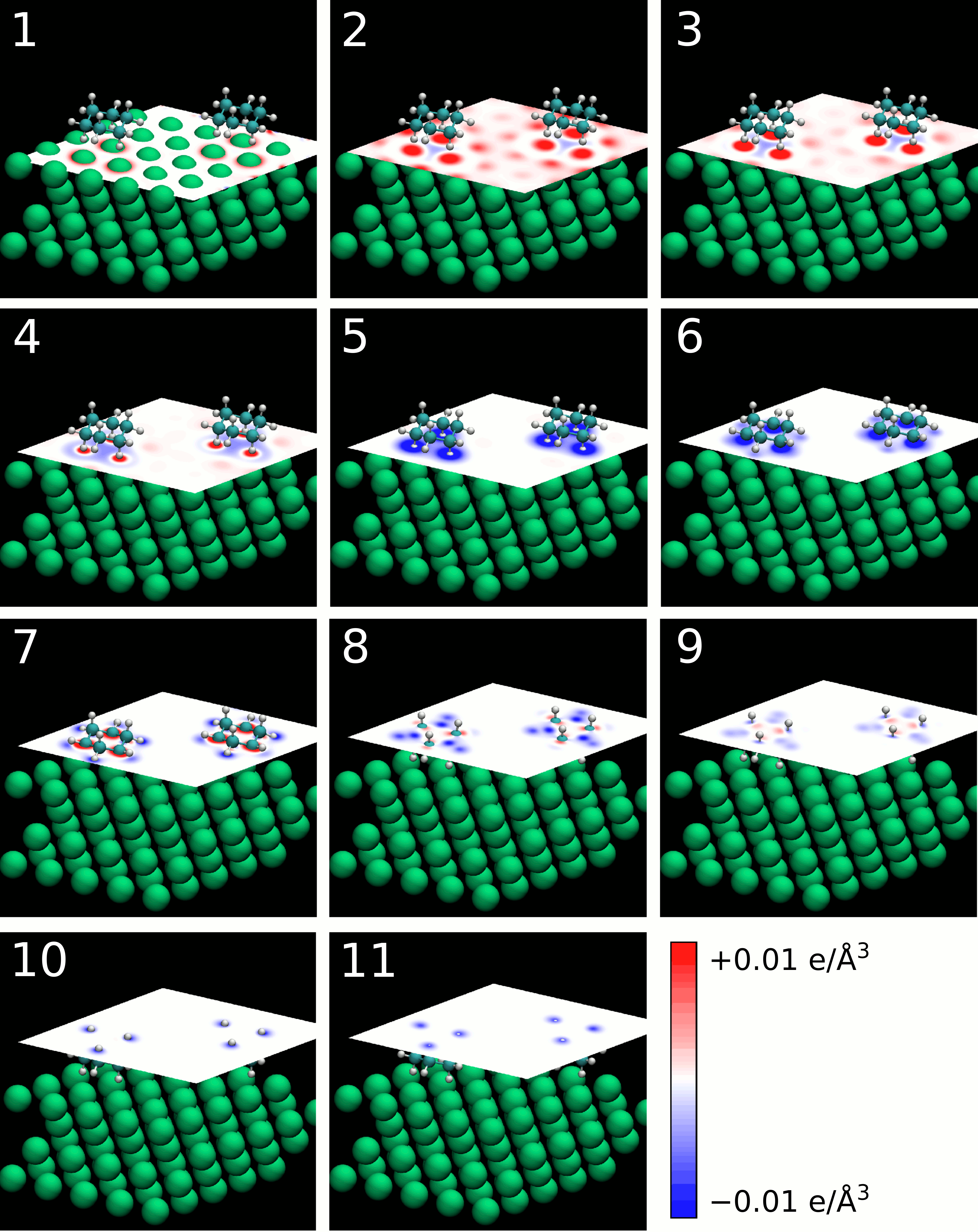}
    \caption{Difference between the electron density of a surface with molecules adsorbed and the sum of isolated surface and isolated molecules, shown at different $x-y$ slices along $z$ coordinate. Red color denotes electron density accumulation, and blue denotes depletion.}
    \label{fig_SI_slices_z}
\end{figure}

Multiple slices of the electron density difference between the full system (surface+molecules) and the superposition of an isolated adsorbate and a clean surface are presented  (figures~\ref{fig_SI_slices_x},~\ref{fig_SI_slices_z}). In these pictures, shades of blue mean electron depletion upon adsorption, and shades of red mean electron accumulation.

%\section{Anharmonicity score}
%\KF{I think I'll just remove NM representation completely. In principle, I could compare H and D, but their NMs are different, so this procedure is ill-defined.

%Alternatively, I can leave only 6 rigid modes, because they should be similar and they are important.}
% The anharmonicity score $\epsilon$ can be calculated for Cartesian components of atomic forces, as shown in the manuscript, and also for forces projected on the normal modes of the adsorbate, as shown in figure~\ref{fig_SI_epsilon_nm}. \KF{Quantum anharmonicity is fairly high, but it cancels out to a large extent when we compare $\rm{C_6H_{12}}$ and $\rm{C_6D_{12}}$.}
% \begin{figure}[htbp]
%     \centering
%     \includegraphics[width=0.4\textwidth]{epsilon-nm.pdf}
%     \caption{Anharmonicity score $\epsilon$ for the vibrational modes of $\rm{C_6H_{12}}$ in PIMD simulation (red) compared to classical-nuclei MD (blue). The modes $m_1^{+}, m_2^{+}$ are coherent stretches of all 6 H atoms pointing towards the surface and towards vacuum, respectively. $m_1^{-}$ is anti-coherent stretch, when 3 surface-pointing hydrogens of one molecule move in counter-phase with those of the second molecule.
%     }
%     \label{fig_SI_epsilon_nm}
% \end{figure}

\section{Estimate of the error of SL-RPC in potential energy}\label{A1}
Our expression 2 in the manuscript %\ref{eq_SLRPC_E_error} 
can be rewritten as 
\begin{equation}
    \frac{k_B T}{2} \sum_\nu \sum_k \left[ 
    \frac{\omega^2_{\rm{mol}}}{\omega_k^2 + \omega^2_{\nu,\rm{full}} + \Delta} - 
    \frac{\omega^2_{\rm{full}}}{\omega_k^2 + \omega^2_{\nu,\rm{full}}} \right],
\end{equation}
where $\Delta = \omega^2_{\nu,\rm{mol}} - \omega^2_{\nu,\rm{full}} $. Since $\Delta$ is much smaller than $\omega_k^2 + \omega^2_{\nu,\rm{full}}$, it can be omitted, and we immediately get eq.~9 from the Ref.~\cite{SL-RPC}. It is a reasonable approximation of one fraction, which gives an error of not more than 10\% in practical cases. However, two fractions have quite close values, therefore the \textit{difference} between them can be even smaller than the error introduced by omitting the $\Delta$. This fact leads to a significant overestimation of the error, if the eq.~9 from~\cite{SL-RPC} is used.

\section{Estimate of the error of SL-RPC in free energy}\label{A2}
Assuming a system to be harmonic, the Hamiltonian  of a ring polymer with $P$ beads in ``physical'' normal modes can be written as
\begin{equation}
    H = K + \sum_{\nu=1}^{3N} \sum_{k=1}^P \left[ \frac{m_{\nu} \omega_P^2}{2}\left(q_\nu^{(k)} - q_\nu^{(k+1)}\right)^2 + \frac{m_{\nu} \omega_\nu^2}{2} q_\nu^{(k)2} \right].
\end{equation}
Here $K$ is a kinetic energy, $\nu$ denotes normal modes (NMs) of a physical system, $m_\nu$ is an effective mass associated with the normal mode $\nu$. Note that here, $k$ stands for a bead index -- in contrast to the following equations, where it will enumerate normal modes of a free ring polymer.

Expanding the spring-terms and rearranging the summation over $k$ (also making use of periodicity of a ring polymer $q^{(P+1)} = q^{(1)}$), one can rewrite a Hamiltonian as following:
\begin{equation}
\begin{split}
    H = K
    + \sum_{\nu=1}^{3N} \sum_{k=1}^P \left[ \frac{m_{\nu} \omega_P^2}{2}\left(2q_{\nu}^{(k)2} - q_{\nu}^{(k)}q_{\nu}^{(k+1)} - q_{\nu}^{(k)}q_{\nu}^{(k-1)}\right) 
    + \frac{m_{\nu} \omega_\nu^2}{2} q_\nu^{(k)2} \right].
\end{split}
\end{equation}
Then, the spring terms can be written in a matrix form (bold below means $P$-dimensional vectors $\{q^j\}, j \in [1,...,P]$ and corresponding square matrices)
\begin{equation}
    V^{spring} = \sum_{\nu=1}^{3N} \frac{m_\nu \omega_P^2}{2} \bm{q_\nu^\top A q_\nu},
\end{equation}
\begin{equation}
    \bm{A} = \begin{bmatrix}
        2 & -1 &  0 & ...&    & -1 \\
       -1 &  2 & -1 &  0 &    &    \\
        0 & -1 &  2 & ...&    &    \\
       ...&  0 & ...&    &    &    \\
          &    &    &    &    & -1 \\
       -1 &    &    &    & -1 &  2 \\
    \end{bmatrix}
\end{equation}
We diagonalize the $\bm{A}$ matrix by performing a normal mode transformation $\bm{C}$:
\begin{equation}
    \bm{A} = \bm{C \tilde{A} C^\top}.
\end{equation}
\begin{equation}
    \bm{\tilde{q}}_\nu = \bm{C q}_\nu
\end{equation}
The matrix $\bm{C}$ is unitary, therefore the transformation to the normal modes of a free ring polymer doesn't change the physical potential term
\begin{equation}
    \frac{m_\nu \omega_\nu^2}{2} \bm{q}_\nu^2 
    = \frac{m_\nu \omega_\nu^2}{2} \bm{\tilde{q}}_\nu^\top C^{-1\top} C^{-1} \bm{\tilde{q}}_\nu 
    = \frac{m_\nu \omega_\nu^2}{2} \bm{\tilde{q}}_\nu^2 .
\end{equation}

The Hamiltonian in the ``double normal-mode'' representation (i.e. the normal modes of a physical system and the normal modes of a free ring polymer) reads as
\begin{equation}
    H = K + \sum_{\nu = 1}^{3N} \sum_{k=0}^{P-1} \frac{m_\nu (\omega_k^2 + \omega_{\nu,\rm{full}}^2)}{2} \tilde{q}_\nu^{(k)2},
\end{equation}
\begin{equation}
    \tilde{q}_\nu^{(k)} = \sum_{j=1}^P C^{P}_{jk} q_\nu^{j}
\end{equation}
where $k$ denotes NMs of a free ring polymer. 
Here and below, ``full'' index stands for the ``expensive'' potential energy surface which describes all interactions in a system, while ``mol'' stands for the ``cheap'' one. In case of spatially-localized contraction, a ``cheap'' potential describes only interactions within an adsorbate.

\textbf{Contraction procedure.}
Given a ring polymer of $P$ beads, one can contract it to a lower dimensionality. Many useful expressions can be found in~\cite{Markland_Manolopoulos_2008}.
Equations 21-22 from \cite{Markland_Manolopoulos_2008}, rewritten in our notation:
\begin{equation}
    q_\nu^{(j')} = \sum_{j=1}^{P'} (T_P^{P'})_{j'j} q_\nu^{(j)},
\end{equation}
where
\begin{equation}
    (T_P^{P'})_{j'j} = \frac{1}{P} \sum_{k=-P'/2}^{P'/2} C_{j'k}^{P'} C_{jk}^{P}
\end{equation}
is a contraction matrix from $(P \times 3N)$ to  $(P' \times 3N)$-dimensional space. It performs transformation $C^P$ to a Fourier space, there it truncates the high-order coefficients and transforms back to a lower-dimensional real space by $C^{P'}$. Similarly, we define an expansion matrix $T_{P'}^P$ from $(P' \times 3N)$ to  $(P \times 3N)$-dimensional space. The expansion procedure is a reverse of a contraction with only difference: instead of truncating Fourier series, we have to expand it from $P'$ to $P$ terms. Since we don't have these coefficients, we set them to be zero.
It can be shown that the potential energies of the $P$- and $P'$-ring polymers are related as
\begin{equation}
    \sum_{k=1}^P V(q_{1}^{(k)},..., q_{3N}^{(k)}) \approx \frac{P}{P'} \sum_{j=1}^{P'} V(q'^{(j)}_1, ..., q'^{(j)}_{3N}),
\end{equation}
with respect to the accuracy of contraction.

The Hamiltonian after the SL-RPC is applied:
\begin{equation}
\begin{split}
    H = K + \sum_{\nu = 1}^{3N}
    \left[ \sum_{k=0}^{P'-1} \frac{m_\nu (\omega_k^2 + \omega_{\nu,\rm{full}}^2)}{2} \tilde{q}_\nu^{(k)2} \right. 
    \left. + \sum_{k=P'}^{P-1} \frac{m_\nu (\omega_k^2 + \omega_{\nu,\rm{mol}}^2)}{2} \tilde{q}_\nu^{(k)2} \right].
\end{split}
\end{equation}
Then, the partition function of this system is
\begin{equation}
    Q = \prod_{\nu=1}^{3N} \left[ \prod_{k=0}^{P'-1}\frac{1}{\beta_P \hbar \sqrt{\omega_k^2 + \omega^2_{\nu,\rm{full}}}} \prod_{k=P'}^{P-1} \frac{1}{\beta_P \hbar \sqrt{\omega_k^2 + \omega^2_{\nu,\rm{mol}}}}  \right]
\end{equation}

The free energy of a (single) physical system:
\begin{equation}
\begin{split}
    F & = - \frac{1}{\beta} \ln(Q) = 
    = - \frac{1}{\beta} \sum_{\nu = 1}^{3N} \left[ - \sum_{k=0}^{P'-1}\ln(\beta_P \hbar \sqrt{\omega_k^2 + \omega^2_{\nu,\rm{full}}})
    - \sum_{k=P'}^{P-1} \ln( \beta_P \hbar \sqrt{\omega_k^2 + \omega^2_{\nu,\rm{mol}}}) \right] = \\
    & = \frac{3NP\ln(\beta_P \hbar)}{\beta} + \frac{1}{2\beta} \sum_{\nu = 1}^{3N} 
    \left[\sum_{k=0}^{P'-1}\ln(\omega_k^2 + \omega^2_{\nu,\rm{full}}) 
    + \sum_{k=P'}^{P-1} \ln(\omega_k^2 + \omega^2_{\nu,\rm{mol}}) \right] .
\end{split}
\end{equation}

The free energy difference between SL-RPC and $P$ beads calculated with full potential:
\begin{equation}
\begin{split}
    \delta F = (F^{RPC} - F^{P\ beads}) 
    = \frac{1}{2\beta} \sum_{\nu = 1}^{3N} \sum_{k=P'}^{P-1} 
    \ln\left(\frac{\omega_k^2 + \omega^2_{\nu,\rm{mol}}}{\omega_k^2 + \omega^2_{\nu,\rm{full}}}\right) 
    = \frac{1}{2\beta} \sum_{\nu = 1}^{3N} \sum_{k=P'}^{P-1} 
    \ln\left(1 + \frac{\omega^2_{\nu,\rm{mol}} - \omega^2_{\nu,\rm{full}}}{\omega_k^2 + \omega^2_{\nu,\rm{full}}} \right).
    \end{split}
\end{equation}

\bibliography{references}